%% file: arxiv_full_version.tex
\documentclass[final]{l4dc2025}

\input{math_commands.tex}

\usepackage{aliascnt}

\usepackage{hyperref}
\usepackage{url}
\usepackage{booktabs}
\usepackage{algorithm}
\usepackage{algorithmic}
\usepackage{nicefrac}
\usepackage{url}
\usepackage{cleveref}
\usepackage{enumitem}

\newtheorem{prbl}{Problem}
\usepackage[font=small]{caption}
\usepackage{times}

\title[Safe PDE Control]{
Safe  PDE Boundary Control with Neural Operators
}
\usepackage{times}

\coltauthor{\Name{Hanjiang Hu} \Email{hanjianghu@cmu.edu}\and
 \Name{Changliu Liu} \Email{cliu6@andrew.cmu.edu}\\
 \addr The Robotics Institute, Carnegie Mellon University}

\begin{document}

\maketitle

\begin{abstract}%
The physical world dynamics are generally governed by underlying partial differential equations (PDEs) with unknown analytical forms in science and engineering problems. Neural network based data-driven approaches have been heavily studied in simulating and solving PDE problems in recent years, but it is still challenging to move forward from understanding to controlling the unknown PDE dynamics. PDE boundary control instantiates a simplified but important problem by only focusing on PDE boundary conditions as the control input and output. However, current model-free PDE controllers cannot ensure the boundary output satisfies some given user-specified safety constraint. To this end, we propose a safety filtering framework to guarantee the boundary output stays within the safe set for current model-free controllers. Specifically, we first introduce a neural boundary control barrier function (BCBF) to ensure the feasibility of the trajectory-wise constraint satisfaction of boundary output. Based on the neural operator modeling the transfer function from boundary control input to output trajectories, we show that the change in the BCBF depends linearly on the change in input boundary, so 
quadratic programming-based safety filtering can be done for pre-trained model-free controllers. Extensive experiments under challenging hyperbolic, parabolic and Navier-Stokes PDE dynamics environments validate the plug-and-play effectiveness of the proposed method by achieving better general performance and boundary constraint satisfaction compared to the vanilla and constrained model-free controller baselines. The code is available at \url{https://github.com/intelligent-control-lab/safe-pde-control}.
\end{abstract}

\begin{keywords}%
PDE boundary control, safe control, learning for control
\end{keywords}

\input{intro}

\input{formulation}

\input{method}

\input{experiment}

\input{conclusion}



\acks{This work is partially supported by the National Science Foundation,
Grant No. 2144489. Funding to attend this conference was provided by the CMU GSA/Provost Conference Funding.}

\bibliography{arxiv_full_version}

\clearpage
\appendix
\section{Related Work}
\label{sec:related}
\paragraph{Control for PDE Dynamics.} PDE control problems can be in-domain control \citep{botteghi2024parametric,zhang2024policy} or boundary control \citep{krstic2008boundary,smyshlyaev2010adaptive}, where the latter is more commonly-seen setting in the real world. As it has been studied for over a decade, backstepping has become a dominant approach for boundary control with known PDE dynamics \citep{krstic2008backstepping,smyshlyaev2004closed}. Recently, learning-based controllers have gotten rid of the requirement of analytical forms of unstable PDE dynamics and become a promising solution to the PDE  control problems \citep{botteghi2024parametric,zhang2024policy,krstic2024neural,qi2023neural,mowlavi2023optimal,wei2024generative,soroco2025pde}.
Regarding the safety of constraint satisfaction in the PDE dynamics, current backstepping-based safe PDE control methods \citep{krstic2006nonovershooting,li2020mean,koga2023safe,wang2023safe} still assume the non-stable PDE dynamics is known. Recently, \cite{hu2025uncertain} introduce safe diffusion models for PDE Control based on conformal prediction to quantify uncertainty.
Instead, we focus on boundary safety constraint satisfiability in the PDE boundary control signal without any prior knowledge of PDE dynamics.







\paragraph{Safe Control with Neural Certificate}
For the control of the ODE dynamical system, there is rich literature regarding learning-based controllers with safety guarantees or certificates \citep{boffi2021learning,dawson2023safe,xiao2023barriernet,lindemann2021learning,chang2019neural,mazouz2022safety}. Neural networks have been used to parameterize the CBFs under complex dynamics with bounded control inputs \citep{liu2023safe, so2023train,zinage2023neural,dawson2022safe,dai2022learning}, which result in forward invariance of the user-specified safe set to guarantee the safety with neural certificate for learning-based controllers \citep{choi2021robust,wei2022persistently,agrawal2021safe,xiao2022sufficient,hsu2023safety}, i.e. once the states enter the safe set, they will never go out. However, forward invariance may not hold in the PDE boundary control setting with commonly-seen highly oscillating trajectories. 
For example, highly-oscillating trajectories may 
go out of the safe set during the early oscillation and break the forward invariance defined by conventional ODE CBFs \citep{liu2014control,ames2014control}, but they could still converge to the constraint satisfaction by the end of time.
Therefore, we focus on boundary feasibility, a new notion introduced in this paper. Approach-wise, the CBF-QP for ODE dynamics  \citep{liu2014control,lindemann2018control,xiao2021high,garg2021robust} does not apply. That is because PDE boundary control does not have Markov property at each control step, due to the infinite-dimensional unobserved non-boundary states. We adopt a neural operator to model the trajectory-to-trajectory mapping and control the change of input boundary through a novel QP formulation. 

\paragraph{Neural Operator Learning for PDEs.} Neural operator learning has become a powerful tool for solving PDEs by learning mappings between function spaces rather than pointwise approximations \citep{kovachki2023neural,brunton2024promising}. Recent research has demonstrated the utility of neural operators in multiple science and engineering fields like fluid dynamics, weather forecasting, and robotics \citep{kochkov2021machine,pathak2022fourcastnet,heiden2021neuralsim,raissi2019physics}. There exist multiple architectures for neural operators based on different mathematical properties of data. \cite{lu2021learning} introduces DeepONet with a branch and a trunk network, and NOMAD \citep{seidman2022nomad} adopts nonlinear decoder map to learn submanifolds in function
spaces, while Green's function-inspired neural operators \cite{li2020fourier, li2020neural, li2020multipole, li2022learning,li2024physics} adopt linear integral kernel representation with various kernel implementations. Learning-based methods \citep{ma2023learning,li2024neuralfluid} are proposed for differentiable simulation of PDE dynamics, but neural control of PDE dynamics is less explored. 
Recent work \citep{manda2024domain} introduces operator learning for mapping from environmental parameters to
the corresponding CBF under HJ-PDE \citep{bansal2021deepreach, chen2024learning}, which does not directly study the PDE control problem.
For the PDE boundary control problem, current works \citep{bhan2023neural,krstic2024neural} only adopt neural operators to learn the integral kernel in backstepping,  which does not release the full potential of neural operator for characterizing and controlling unknown dynamics. 
The proposed work is the first to leverage neural operators to learn the direct mapping from control input to boundary output as a transfer function.

\section{Proofs}
\subsection{Preliminary}
\begin{definition}[Boundary Feasibility for Finite-time Constraint Satisfaction] (restated from Definition \ref{def:pf})
\label{def:pf_app}
    With state $u(x,t)$ subjected to closed-loop PDE dynamics in \Cref{eq:pde_dynamics} with the boundary control input $U(t)$,  the boundary control output $Y(t)$ is defined to be feasible over $\gT$ within the given user-specified safe set $\gS_0\in\gS$ if the following holds,
    \begin{align}
    \label{eq:pf_app}
        \exists t_0\in\gT, \forall  t_0 \leq t\leq T, Y(t):=u(\vzero,t)\in \gS_0, \text{ where } u(\vone,t)=U(t), u(x,0)\equiv U(0).
    \end{align}
\end{definition}

\begin{definition}[Neural operator for input-output boundary mapping] 
\label{def:NO_formulate}
 Neural operator from \Cref{sec:NO_formulate} $\gG_\theta:\{U: \gT\rightarrow\gS\}\mapsto\{Y: \gT\rightarrow\gS\}$ can be formalized as
\begin{align}
    &Y(t)=\gG_\theta(U)(t)=Q(v_L(t)), v_0(t) =P(U(t)), \text{ where each layer $v_l(t)$ is}\\
    &v_{l+1}(t) = \gI_l(v_l)(t) = \sigma_{l+1}\left(W_lv_l(t)+\int_\gT\kappa^{(l)}(t,s)v_l(s)ds+b_l(t)\right), l=0,1,\dots,L-1
\end{align}
where  $\sigma_{l+1}:\sR^{d_{v_{l+1}}}\rightarrow\sR^{d_{v_{l+1}}}$ is the activation function, $W_l\in\sR^{d_{v_{l+1}}\times d_{v_l}}$ is the local linear operator, $P\in \sR^{v_0\times\text{dim}(\gS)}$ and $Q\in \sR^{\text{dim}(\gS)\times v_L}$ are lifting and projection matrix, $\kappa^{(l)}\in C(\gT\times\gT;\sR^{d_{v_{l+1}}\times d_{v_l}})$ is the  kernel function for integration, and $b_l\in C(\gT;\sR^{d_{v_{l+1}}})$ is the bias function. And $P,Q,W_l,\kappa^{(l)},b_l,l=0,1,\dots,L-1$ are  parameterized with neural networks $\theta$.
\end{definition}

\subsection{Proof of Theorem \ref{thm:pf_thm}}
\label{sec:proof_pf}
\begin{theorem}[Boundary Feasibility with Boundary Control Barrier Function]
\label{thm:pf_thm_app}
For the state $u(x,t)$ from the closed-loop PDE dynamics with boundary control input $U(t)=u(\vone,t), u(x,0)\equiv U_0$, the boundary feasibility of boundary  output $Y(t)=u(\vzero,t)$ over $\gT=[0,T]$ within user-specified safe set $\gS_0$ is guaranteed with neural BCBF $\phi(t,Y)$ if the following holds $\forall t\in\gT$
\begin{align}
\label{eq:cbf_pf_app}
    \left(\gS_{\phi,t}:=\{Y\mid \phi(t,Y)\leq 0\}\subseteq \gS_0\right)\bigwedge \left(\partial_Y\phi\cdot \frac{d Y}{d t} +\partial_t\phi+ \alpha \phi(t,Y) + C_{\alpha,T}\phi(0,U_0) \leq 0\right)
\end{align}
where $C_{\alpha,T} := \frac{\alpha}{e^{\alpha T}-1}>0$ is a constant for  finite-time convergence. Similarly, the boundary feasibility 
  with neural BCBF $\phi(Y)$ holds if \Cref{eq:cbf_pf} holds by letting  $\partial_Y\phi=\nabla_Y\phi, \partial_t\phi=0$.
\end{theorem}
\begin{proof}
    To show the boundary feasibility of the boundary output of $Y(t)$ within user-specified safe set $\gS_0$, by Definition \ref{def:pf_app}, we need to show
    \begin{align}
        \exists t_0 \in [0,T], s.t. \forall t\in[t_0, T], Y(t)\in\gS_0.
    \end{align}
    With the sublevel set  $\gS_{\phi,t}$ being the subset of $\gS_0$, i.e., $\gS_{\phi,t}:=\{Y\mid \phi(t,Y)\leq 0\}\subseteq \gS_0$, it is sufficient to prove 
    \begin{align}
    \label{eq:sufficient_condition}
         \exists t_0 \in [0,T], s.t. \forall t\in[t_0, T], \phi(t,Y(t)) \leq 0.
    \end{align}
    Now denote $\psi(t):= \phi(t,Y(t))$, by initial constant boundary condition $Y(0)=u(\vzero,0)=u(\vone,0)=U_0$, we have the following equivalent inequalities hold,
    \begin{align}
        \partial_Y\phi\cdot \frac{d Y}{d t} +\partial_t\phi+ \alpha \phi(t,Y) + C_{\alpha,T}\phi(0,Y(0)) &\leq 0
        \\
        \Longleftrightarrow \frac{d\phi(t,Y(t))}{dt}+ \alpha \phi(t,Y) + C_{\alpha,T}\phi(0,Y(0)) &\leq 0\\
        \Longleftrightarrow \frac{d\psi(t)}{dt}+ \alpha \psi(t) + C_{\alpha,T}\psi(0) &\leq 0\\
        \Longleftrightarrow e^{\alpha t}\frac{d\psi(t)}{dt}+ e^{\alpha t}\alpha \psi(t) + e^{\alpha t}C_{\alpha,T}\psi(0) &\leq 0, \forall t\in [0,T]
\\
        \Longleftrightarrow \frac{d(e^{\alpha t}\psi(t)+\frac{C_{\alpha,T}\psi(0)}{\alpha}e^{\alpha t})}{dt} &\leq 0
     \end{align}
     So we have the function $e^{\alpha t}\psi(t)+\frac{C_{\alpha,T}\psi(0)}{\alpha}e^{\alpha t}$ be non-increasing over $t\in[0,T]$. By $T>0$, we have
     \begin{align}
         [e^{\alpha t}\psi(t)+\frac{C_{\alpha,T}\psi(0)}{\alpha}e^{\alpha t}]|_{t=T} &< [e^{\alpha t}\psi(t)+\frac{C_{\alpha,T}\psi(0)}{\alpha}e^{\alpha t}]|_{t=0}\\
         \Longleftrightarrow e^{\alpha T}\psi(T)+\frac{e^{\alpha T}}{e^{\alpha T}-1}\psi(0) & <\psi(0) + \frac{1}{e^{\alpha T}-1}\psi(0) \\
\Longleftrightarrow e^{\alpha T}\psi(T) & < 0 \\
\Longleftrightarrow \psi(T) & < 0 \\
\Longleftrightarrow \phi(T, Y(T)) & < 0 
     \end{align}
     So at least at $t_0=T$, $\phi(t_0, Y(t_0)) < 0$, which proves \Cref{eq:sufficient_condition} holds and the original theorem has been proved.
     Furthermore, let us look at the boundary feasible steps. Since $e^{\alpha t}\psi(t)+\frac{C_{\alpha,T}\psi(0)}{\alpha}e^{\alpha t} = e^{\alpha t}(\psi(t)+\frac{C_{\alpha,T}\psi(0)}{\alpha})$ is non-increasing, with the strictly increasing and positive $e^{\alpha t}$, it is easy to find function $\psi(t)+\frac{C_{\alpha,T}\psi(0)}{\alpha}$ being non-increasing, i.e. $\psi(t)$ is non-increasing. Therefore, if $U_0 \leq0$, $\phi(t, Y(t))<\phi(0, Y(0))=U_0<0, \forall t\in[0,T]$. If $U_0 > 0$, since MLP-ReLU parameterized neural BCBF $\phi$ and boundary control output $Y$ are continuous, by mean value theorem, we have
     \begin{align}
         \phi(0, Y(0)) >0, \phi(T, Y(T)) <0\Rightarrow \exists t_0\in[0,T], \phi(t_0, Y(t_0)) = 0.
     \end{align}
     Since $\psi(t)=\phi(t, Y(t)) $ is non-increasing, we have
     \begin{align}
         \exists t_0 \in [0,T], s.t. \forall t\in[t_0, T], \phi(t,Y(t)) \leq 0,
     \end{align}
     which concludes the proof.
\end{proof}

\subsection{Proof of Theorem \ref{thm:certificate}}
\label{sec:proof_cert}
\begin{theorem}[Boundary Feasibility with Neural Operator]
\label{thm:certificate_app}
Assuming the neural operator $\gG_\theta$ as an exact map from boundary input $U(t)$ to output $Y(t)$
for  an unknown closed-loop PDE dynamics  without model mismatch, the boundary control input $U(t)$  is guaranteed to induce  boundary feasibility of output  $Y(t)$ over $\gT=[0,T]$ within the sublevel set of neural BCBF $\phi$  if $ U(t)$ satisfies 
\begin{align}
\label{eq:pf_G_app}
    \partial_Y\phi(t,\gG_\theta(U)) \frac{d \gG_\theta(U)(t)}{d t} +\partial_t\phi(t,\gG_\theta(U))+ \alpha \phi(t,\gG_\theta(U)) + C_{\alpha,T}\phi(0, U(0)) \leq 0, \forall t\in\gT
\end{align}
 where $C_{\alpha,T} = \frac{\alpha}{e^{\alpha T}-1}$, and $\frac{d \gG_\theta(U)(t)}{d t}$ can be  found below with $\prod_1^0 (\cdot):=1$,
\begin{align}
\label{eq:de_no_full_app}
&\frac{d \gG_\theta(U)(t)}{d t} = \nabla Q^\top\prod_{l=0}^{L-1}\left(\text{Diag}(\sigma_{L-l}')W_{L-1-l}\right)\nabla P^\top \frac{dU(t)}{dt} + \nabla Q^\top\text{Diag}(\sigma_L') \sum_{i=0}^{L-1} \left(\big[\prod_{j=1}^i W_{L-j}\right. \notag\\&\left. \text{Diag}(\sigma_{L-j}')\big] \left(\int_\gT\frac{\partial \kappa^{(L-1-i)}(t,s)}{\partial t}v_{L-1-i}(s)ds+\frac{db_{L-1-i}(t)}{dt}\right)\right) = \Lambda_\theta(t) \dot U(t) + \mu_\theta(t)
\end{align}
\end{theorem}
\begin{proof} To show the boundary feasibility over sublevel set of $\phi$ hold, we first want to show \Cref{eq:de_no_full_app} holds.
According to Definition \ref{def:NO_formulate}, we first rewrite the neural operator  as
\begin{align}
\label{eq:no_forward_app_pf}
    &Y(t)=\gG_\theta(U)(t)=Q(v_L(t)), v_0(t) =P(U(t)), \text{ where each layer $v_l(t)$ is}\notag\\
    &v_{l+1}(t) = \gI_l(v_l)(t) = \sigma_{l+1}\left(W_lv_l(t)+\int_\gT\kappa^{(l)}(t,s)v_l(s)ds+b_l(t)\right), l=0,1,\dots,L-1
\end{align}
where $P,Q,W_l,\kappa^{(l)},b_l,l=0,1,\dots,L-1$ are neural networks,  kernel function $\kappa^{(l)}$,  activation function $\sigma_l$ and bias function $b_l$ are first-order differential.  Since the operator shares the same input function domain and output function domain over $t\in\sR^+$, applying chain rule to \Cref{eq:no_forward_app_pf}, we can find the derivative with respect to $t$ for each layer as,
\begin{align}
    \label{eq:no_backward_app}
    &\frac{dY(t)}{dt} = \nabla Q^\top\frac{dv_L(t)}{dt},   \frac{v_0(t)}{dt} = \nabla P^\top \frac{dU(t)}{dt}, \text{ for each  derivative $\frac{dv_{l+1}(t)}{dt}$ } l = L-1, \dots,0, \\
    \label{eq:de_int_layer_app}
    &\frac{dv_{l+1}(t)}{dt} = \gJ_l(\frac{dv_l}{dt})(t) = \text{Diag}(\sigma_{l+1}')\left(W_l\frac{dv_l(t)}{dt}+\int_\gT\frac{\partial \kappa^{(l)}(t,s)}{\partial t}v_l(s)ds+\frac{db_l(t)}{dt}\right)
\end{align}
Now put \Cref{eq:de_int_layer_app} into \Cref{eq:no_backward_app} recursively, we have
\begin{align}
     &\frac{d\gG(U)(t)}{dt} =\nabla Q^\top\frac{dv_L(t)}{dt} \\
     =&\nabla Q^\top\text{Diag}(\sigma_{L}')W_{L-1}\frac{dv_{L-1}(t)}{dt}+\nabla Q^\top\text{Diag}(\sigma_{L}')\left(\int_\gT\frac{\partial \kappa^{(L-1)}(t,s)}{\partial t}v_{L-1}(s)ds+\frac{db_{L-1}(t)}{dt}\right)  \\
     =& \nabla Q^\top\text{Diag}(\sigma_{L}')W_{L-1}\text{Diag}(\sigma_{L-1}')W_{L-2}\frac{dv_{L-2}(t)}{dt} + \nabla Q^\top\text{Diag}(\sigma_{L}')W_{L-1}\cdot\text{Diag}(\sigma_{L-1}')\cdot\notag\\
     &\left(\int_\gT\frac{\partial \kappa^{(L-2)}(t,s)}{\partial t}v_{L-2}(s)ds+\frac{db_{L-2}(t)}{dt}\right)+ \nabla Q^\top\text{Diag}(\sigma_{L}')(\int_\gT\frac{\partial \kappa^{(L-1)}(t,s)}{\partial t}v_{L-1}(s)ds\notag\\
     &+\frac{db_{L-1}(t)}{dt}) \\
     =& \dots \text{(recursively apply \Cref{eq:de_int_layer_app})}\notag\\
     =&\nabla Q^\top\text{Diag}(\sigma_{L}')W_{L-1}\dots\text{Diag}(\sigma_{1}')W_{0}\frac{dv_{0}(t)}{dt}+ \nabla Q^\top\text{Diag}(\sigma_{L}')W_{L-1}\text{Diag}(\sigma_{L-1}')\cdots W_{1}\notag\\
     &\text{Diag}(\sigma_{1}')\left(\int_\gT\frac{\partial \kappa^{(0)}(t,s)}{\partial t}v_{0}(s)ds+\frac{db_{0}(t)}{dt}\right)  + \dots +\nabla Q^\top\text{Diag}(\sigma_{L}')W_{L-1}\cdot\text{Diag}(\sigma_{L-1}')\cdot\notag\\
     &\left(\int_\gT\frac{\partial \kappa^{(L-2)}(t,s)}{\partial t}v_{L-2}(s)ds+\frac{db_{L-2}(t)}{dt}\right)+ \nabla Q^\top\text{Diag}(\sigma_{L}')(\int_\gT\frac{\partial \kappa^{(L-1)}(t,s)}{\partial t}v_{L-1}(s)ds \notag\\
     &\frac{db_{L-1}(t)}{dt})
     \\
     \label{eq:final_derive}
     =&\nabla Q^\top\prod_{l=0}^{L-1}\left(\text{Diag}(\sigma_{L-l}')W_{L-1-l}\right)\nabla P^\top \frac{dU(t)}{dt} + \nabla Q^\top\text{Diag}(\sigma_L') \sum_{i=0}^{L-1} \left(\big[\prod_{j=1}^i W_{L-j} \text{Diag}(\sigma_{L-j}')\big]\cdot \right. \notag\\&\left.\left(\int_\gT\frac{\partial \kappa^{(L-1-i)}(t,s)}{\partial t}v_{L-1-i}(s)ds+\frac{db_{L-1-i}(t)}{dt}\right)\right) 
\end{align}
Note that the final expression in \Cref{eq:final_derive} is actually linear with respect to $\dot U(t)$ and the weight and bias terms only depend on the parameters of the neural operator $\theta$ and the values at time $t$. Denote the linear weight and bias as $\Lambda_\theta(t),\mu_\theta(t)$
\begin{align}
    &\Lambda_\theta(t) := \nabla Q^\top\prod_{l=0}^{L-1}\left(\text{Diag}(\sigma_{L-l}')W_{L-1-l}\right)\nabla P^\top, 
    \mu_\theta(t) := \nabla Q^\top\text{Diag}(\sigma_L') \cdot\\&\sum_{i=0}^{L-1} \left(\big[\prod_{j=1}^i W_{L-j} \text{Diag}(\sigma_{L-j}')\big]\cdot \left(\int_\gT\frac{\partial \kappa^{(L-1-i)}(t,s)}{\partial t}v_{L-1-i}(s)ds+\frac{db_{L-1-i}(t)}{dt}\right)\right),
\end{align}
then we have $$\frac{dY(t)}{dt} = \frac{d\gG(U)(t)}{dt}=\Lambda_\theta(t) \dot U(t) + \mu_\theta(t).$$
Since $Y(t) = \gG(U)(t)$, \Cref{eq:pf_G_app} is equivalent to $$\partial_Y\phi\cdot \frac{d Y}{d t} +\partial_t\phi+ \alpha \phi(t,Y) + C_{\alpha,T}\phi(0,U(0)) \leq 0.$$
Similar to the proof of Theorem \ref{thm:pf_thm_app}, we have
\begin{align}
         \exists t_0 \in [0,T], s.t. \forall t\in[t_0, T], \phi(t,Y(t)) \leq 0,
    \end{align}
which concludes the proof of  boundary feasibility over the sublevel set of $\phi$.
\end{proof}

\section{Experiment Details}
\label{sec:exp_app}
\subsection{Experiment Setting}
\label{sec:exp_setup_app}
\paragraph{Data preparation.} For 1D environments, the boundary input is $U(t) = u(1,t)$ while the boundary output for the hyperbolic PDE is $Y(t) = u(0,t)$ and the  boundary output for the parabolic PDE $Y(t) = u(0.5,t)$ since $u(0,t)\equiv0$. For the 2D environment, the boundary input is the x-axis consistent boundary condition, i.e., $u(x, 1, t)\equiv U(x), v(x, 1, t)\equiv0,\forall x\in[0,1]$. The boundary output is $Y(t)=u(0.5, 0.95, t), v(x, 0.95, t)\equiv0,\forall x\in[0,1]$, which has the maximum speed except for control input and can be viewed as an indicator for  tracking  performance.  Note that we focus on the boundary output, which only depends on time in high-dimensional cases.  The temporal resolution of collected trajectories is consistent with the control frequency of each environment in \cite{bhan2024pde}, i.e., 50 steps in 5s for hyperbolic PDE, 1000 steps in 1s for parabolic PDE and 200 steps in 0.2s for Navier-Stokes PDE.
We train the RL models PPO and SAC following the default hyper-parameters and unstable PDE settings in \cite{bhan2024pde} for hyperbolic and parabolic equations, while directly adopting the pre-trained models under default Navier-Stokes equation. For the data collection in the 1D hyperbolic equation, we evaluate the backstepping-based model \citep{krstic2008backstepping}, PPO and SAC models with random initial conditions $U_0\in[1,10]$ and collect 50k pairs of input and output $u(1,t), u(0,t)$ trajectories for each model. Similarly, for the 1D parabolic equation, we evaluate the backstepping-based model \citep{smyshlyaev2004closed}, PPO and SAC models with random initial conditions $U_0\in[1,10]$ and collect 50k pairs of input and output $u(1,t), u(0.5,t)$ trajectories for each model.  For the Navier-Stokes equation, we evaluate the model-based optimization method \citep{pyta2015optimal}, PPO and SAC models with random initial conditions $u_0\in[-0.1,0.1]$ and default tracking ground truth and collect 10k pairs of input and output $u(0.05,1,t), u(0.5,0.95,t)$ trajectories for each model. After the data pairs are collected, we annotate the safety label with pre-defined safe constraints based on the original performance of each policy. We specify one-sided  safe sets $\gS_0=\{Y:AY<b\}$ for stabilization tasks and two-sided safe sets $\gS_0=\{Y:|Y-Y_{gt}|<b\}$  for tracking tasks.
Specifically, for the hyperbolic equation, $Y<1$ for PPO and $Y<0$ for SAC; for the parabolic equation, $Y<0.6$ for PPO and $Y>-0.26$ for SAC; for the Navier-Stokes equation, $|Y-Y_{gt}|<0.145$ for PPO and SAC models. Then we randomly split $90\%$ as a training dataset and leave others as a test set. The safe reinforcement learning baselines are based on cumulative costs with empirical performance towards the safety constraints of the output boundary  \citep{achiam2017constrained, ha2020learning,liu2022constrained,liu2024datasets}.

\paragraph{Model training.} 
To train the neural operator models, we adopt the public package \citep{NeuralOperators}, using the default gelu-activation model of FNO with channels of $(2, 64, 64, 64, 64, 64, 128, 1)$ and $16$ modes,  MNO with channels of $(2, 64, 64, 64, 64, 64, 1)$ and $16$ modes. All the models are trained for 100 epochs with learning rate $10^{-3}$, $\ell$-2 regularization weight is $10^{-4}$, ADAM optimizer and $\ell$-2 loss. The resolutions and scales of hyperbolic, parabolic, and Navier-Stokes trajectories are 50, 1000, and 200 for 5s, 1s, and 0.2s, respectively. We keep the same setting for different environments and remark that we do not fully exploit the potential for the best performance of neural operators since it is not the main focus of this work.
For the neural BCBF training, we directly use the finite difference of $Y(t)$ collected from real PDE dynamics instead of the gradient of the neural operator to avoid noise. Following the implementation of \cite{dawson2022safe,zhang2023exact,hu2024verification}, we adopt 4-layer MLPs with ReLU with layer dimensions of (16,64,16,1) to model neural BCBFs. The time $t$ is concatenated with $Y(t)$ as input for time-dependent neural BCBF $\phi(t, Y)$ while only $Y(t)$ is input for time-independent neural BCBF $\phi(t, Y)$. To construct the safe set loss in \Cref{eq:loss_S}, we adopt all the sampled steps along trajectories with unsafe labels while only choosing the "latest" safe sampled steps where boundary feasibility is satisfied in Definition \ref{def:pf}, i.e. once $Y(t)$ is with safe label, it will never become unsafe in finite time $T$. For the boundary feasibility loss in \Cref{eq:total_loss}, due to too much data close to 0, we adopt a random drop of close-to-0 data to balance the output boundary data distribution. Specifically, for the hyperbolic equation, we keep 20\% data within [-0.1,0.1] while keeping 20\% data within [-0.01,0.01] for the parabolic equation. Following \cite{liu2023safe}, we adopt a regularization loss to avoid the shrinking of the sublevel set during training with a default weight of 1. We train all models with ADAM for 20 epochs with an initial learning rate of 0.01. The learning rate decay rate is 0.2 after each 4 epochs. The code can be found at \url{https://github.com/intelligent-control-lab/safe-pde-control}.

\paragraph{Discrete-time Implementation.} 
We remark that iterative filtering with the prediction of $Y(t)$ at each step aims to avoid large approximation errors in \Cref{eq:de_no_full} in the discrete-time setting compared to one-time filtering for the whole trajectory. Besides, 
as the computation of QP is not yet real-time, it is not yet ready to interact with the real PDE dynamics. 
we adopt the predicted $Y(t)$ from the neural operator after each filtering step instead of real PDE dynamics. To handle the model mismatch issue between neural operator modeling and real underlying PDE dynamics, the filtering threshold $\eta >0$ is introduced as a workaround and we leave the study of model mismatch of PDE dynamics as future work. Specifically, the safety filter is disabled when $\eta =0$. The larger $\eta$ is, the more boundary feasibility within the safe set will be achieved, showing a trade-off between stabilization and constraint satisfaction.
The final control trajectory is found through \Cref{eq:filter_threshold} with threshold $\eta=2$ as default, mitigating the discrepancy between the PDE environment and the neural operator.

\begin{align}
\label{eq:filter_threshold}
    U_{\text{safe}}(t) = \int_{0}^t \dot U(\tau) d\tau + U_{\text{nominal}}(0),  \dot U(\tau) = \begin{cases}
        \dot U_{\text{safe}}(\tau), \text{if } \|\dot U_{\text{safe}}(\tau) - \dot U_{\text{nominal}}(\tau)\| \leq  \eta,\\
        \dot U_{\text{nominal}}(\tau),  \text{ otherwise.}
    \end{cases}
\end{align}
We remark that iterative filtering with the prediction of $Y(t)$ at each step aims to avoid large approximation errors in \Cref{eq:de_no_full} in the discrete-time setting compared to one-time filtering for the whole trajectory. 

\subsection{Additional Results}
\label{sec:more_threshold}


\begin{table}[t]
\vspace{-3mm}
\caption{Comparison time-independent and time-dependent safety filtering in different equations.}
\vspace{-5mm}
\begin{center}
\label{tab:ablation_different_BCBF_app}
\resizebox{0.9\textwidth}{!}{
    \begin{tabular}{cccc} 
 \toprule
1D parabolic equation                        & \begin{tabular}[c]{@{}c@{}}Reward (mean$\pm$std)\\ (starting at $\sim$0)\end{tabular}  & \begin{tabular}[c]{@{}c@{}}Feasible Rate\\ (100 episodes)\end{tabular}  &  \begin{tabular}[c]{@{}c@{}}Average Feasible Steps\\ ( 1000 control steps)\end{tabular} \\ \midrule
PPO with filtering  of $\phi(Y)$ & 162.9$\pm$19.6  & 0.46                        & \textbf{519.4}    \\
PPO with filtering  of $\phi(t,Y)$  & {168.2}$\pm$23.5  & \textbf{0.81}                        & \textbf{507.0}                                                        \\\midrule
SAC with filtering  of $\phi(Y)$ & {157.9}$\pm$6.9   & \textbf{0.92}                        &   \textbf{543.2}   \\
PPO with filtering  of $\phi(t,Y)$ & \textbf{157.5}$\pm$6.8   & \textbf{0.87}                        &  \textbf{449.8}     \\
\bottomrule
\toprule
2D Navier-Stokes equation                       & \begin{tabular}[c]{@{}c@{}}Reward (mean$\pm$std)\\ (starting at $\sim$-100)\end{tabular}  & \begin{tabular}[c]{@{}c@{}}Feasible Rate\\ (100 episodes)\end{tabular}  &  \begin{tabular}[c]{@{}c@{}}Average Feasible Steps\\ ( 200 control steps)\end{tabular} \\ \midrule
 PPO with filtering  of $\phi(Y)$ & {-5.37}$\pm$0.01  & 0.86                       & 2.2   \\
 PPO with filtering of $\phi(t,Y)$ & -5.72$\pm$0.17  & \textbf{0.99}                       & \textbf{32.0}                                                          \\\midrule
SAC with filtering  of $\phi(Y)$ & {-18.05}$\pm$1.14  & 0.79                        & 17.8      \\
 SAC with filtering of $\phi(t,Y)$ & -18.36$\pm$1.25  & \textbf{0.85}                        & \textbf{21.3}       \\
\bottomrule
\end{tabular}
}
\end{center}
\vspace{-7mm}
\end{table}

\begin{figure}
    \centering
    \includegraphics[width=0.45\linewidth]{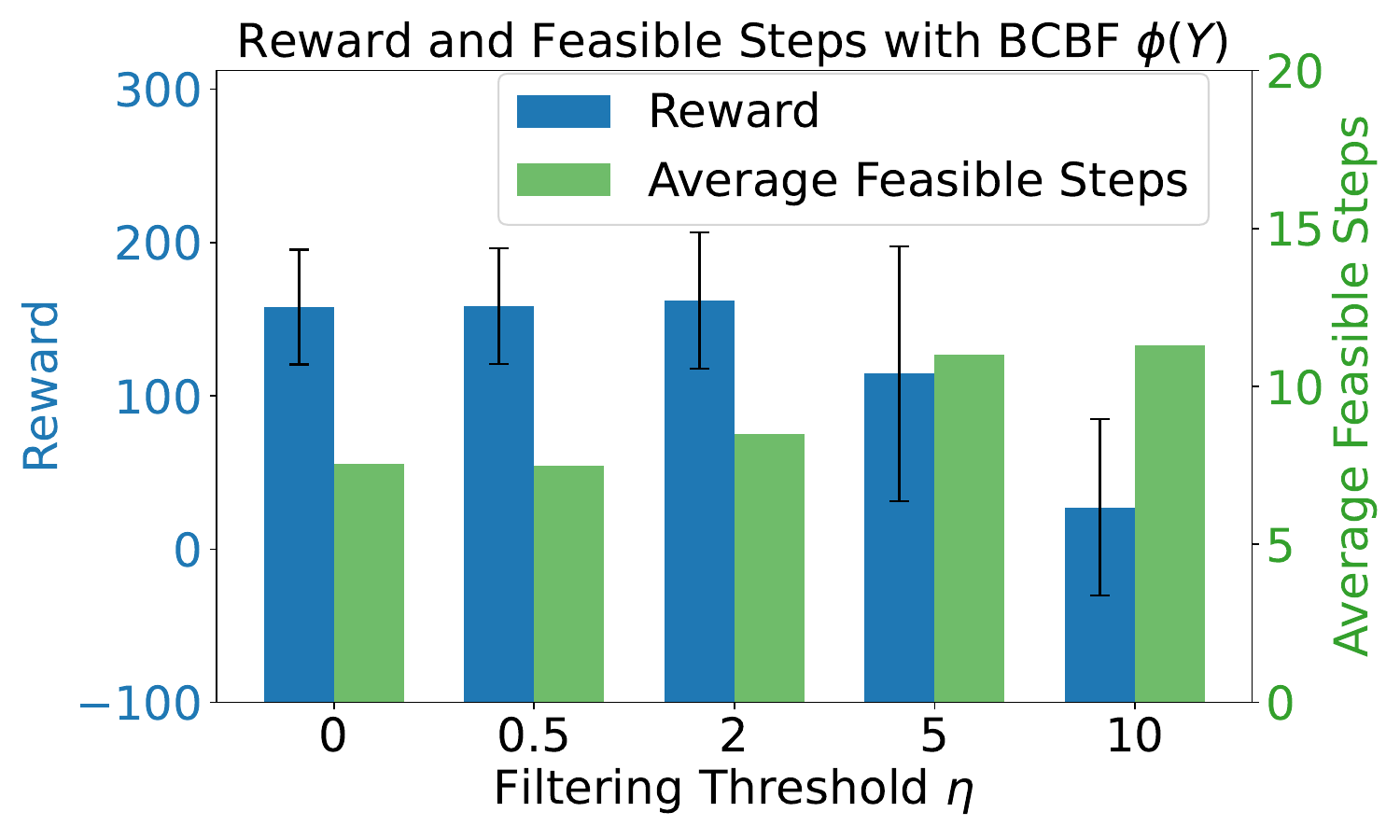}\includegraphics[width=0.45\linewidth]{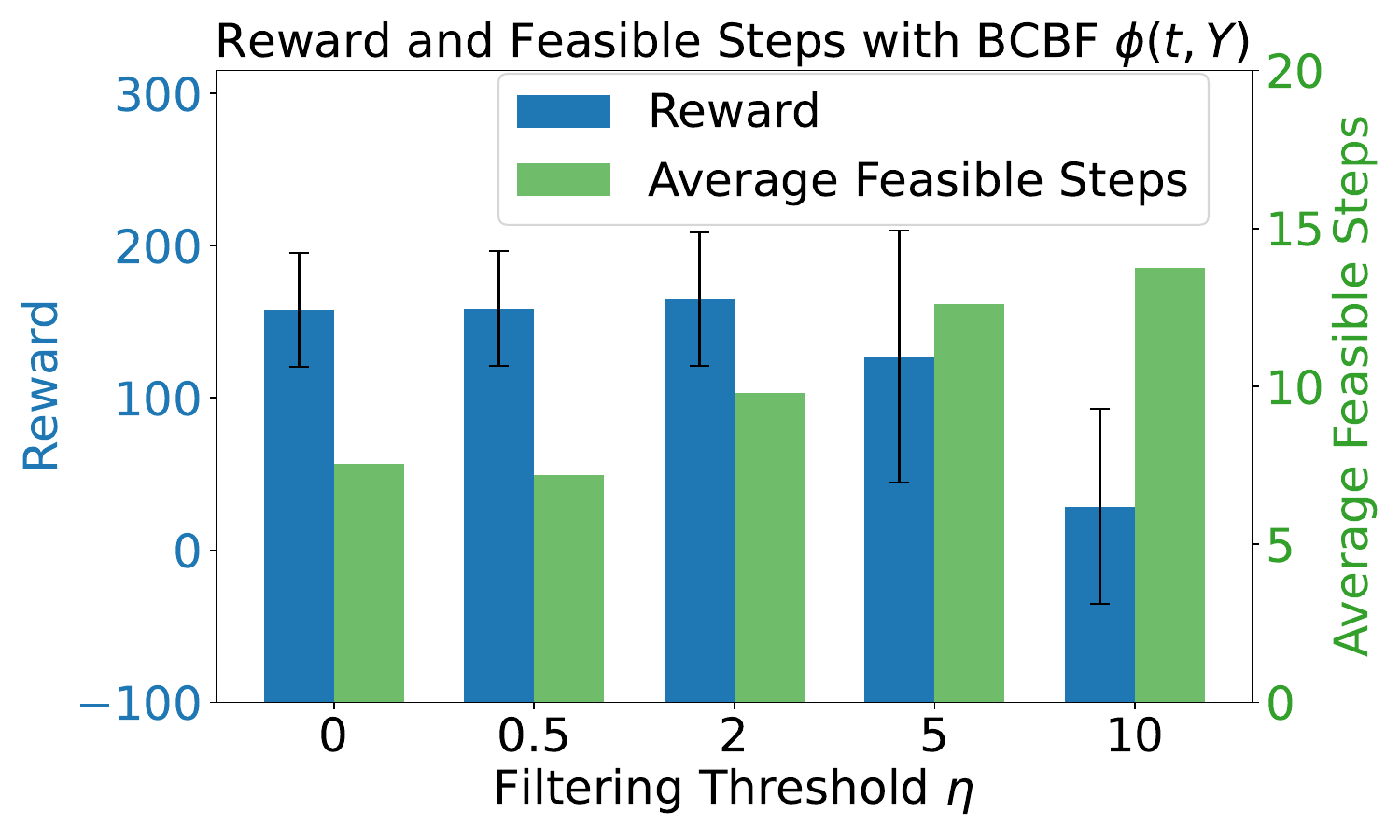}
    \vspace{-4mm}
    \caption{The reward and feasible rate under different filtering threshold $\eta$ in \Cref{eq:filter_threshold} with BCBF $\phi(Y)$ (left) and $\phi(t,Y)$ (right) for PPO model in hyperbolic equation. Note that $\eta=0$ indicates the vanilla PPO model without safety filtering.}
    \vspace{-5mm}
    \label{fig:threshold}
\end{figure}

\paragraph{Influence of filtering threshold.} 
Since the boundary output $Y(t)$ is predicted from the neural operator in \Cref{alg}, the model mismatch significantly influences the performance of safety filtering, where we handle it through a filtering threshold $\eta$ in \Cref{eq:filter_threshold}. We investigate it to show the trade-off between general performance and boundary feasibility. From \Cref{fig:threshold}, it can be seen that as the threshold goes up, the reward first slightly increases and then drops significantly, showing that the strong safety filtering may hurt the stability of the PPO controller due to the model mismatch between direct boundary mapping with the neural operator and underlying PDE dynamics. Besides, with a larger filtering threshold $\eta$, the average feasible steps become larger as the safety filtering becomes stronger, especially for time-dependent BCBF $\phi(t,Y)$, guaranteeing constraint satisfaction over boundary output. With small $\eta$, the average feasible steps may be less than the one without filtering because of more feasible trajectories with last-step feasibility. Safety filtering aligns with the stabilization to increase the reward, but the noise from the model mismatch between the neural operator and real dynamics will make the performance collapse if the safety filtering is too strong.  For the boundary feasibility, we can see that the average feasible steps keep going up as the threshold increases, showing that the finite-time convergence is enhanced for the feasible trajectories. 
However, when the threshold becomes too large, e.g. $\eta=10$, the feasible rate also decreases significantly because the system is no longer stable, as the reward indicates. 

\paragraph{Comparison of asymptotic and finite-time boundary feasibility.} In \Cref{tab:finitetime_PF}, we show the comparison of safety filtering with BCBF $\phi(t,Y)$ for 1D hyperbolic equation for asymptotic and finite-time boundary feasibility. Asymptotic boundary feasibility is with the neural BCBF trained and tested with $C_{\alpha,T}=\lim_{T\to\infty}\frac{\alpha}{e^{\alpha T}-1}=0$ while finite-time boundary feasibility is with $C_{\alpha,T}=0.02$ using $T=50$. It can be seen that BCBF with finite-time feasibility has a better feasible rate, especially the SAC model, as asymptotic feasibility is weaker than finite-time feasibility and takes longer steps to converge. However, for the general reward performance, since asymptotic feasibility causes weaker filtering effects, the reward tends to be closer to the vanilla reward without filtering compared to finite-time feasibility, which is validated in \Cref{tab:finitetime_PF}.
\begin{table}[]
\caption{Results of filtering with BCBF $\phi(t,Y)$ for 1D hyperbolic equation for  asymptotic $C_{\alpha,T}=\lim_{T\to\infty}\frac{\alpha}{e^{\alpha T}-1}=0$ and finite-time  $C_{\alpha,T}=\frac{\alpha}{e^{\alpha T}-1}=0.02$ at $T=50,\alpha=10^{-5}$.}
\begin{center}
\label{tab:finitetime_PF}
\resizebox{0.9\textwidth}{!}{
    \begin{tabular}{cccc} \toprule
Different neural operators                    & \begin{tabular}[c]{@{}c@{}}Reward (mean$\pm$std)\\ (starting at $\sim$-300)\end{tabular}  & \begin{tabular}[c]{@{}c@{}}Feasible Rate\\ (100 episodes)\end{tabular}  &  \begin{tabular}[c]{@{}c@{}}Average Feasible Steps\\ ( 50 control steps)\end{tabular}  \\ \midrule
PPO for asymptotic feasibility    & 163.8$\pm$40.6  & 0.70                        & 8.1                                                         \\
PPO for finite-time feasibility    & \textbf{165.0}$\pm$43.7  & \textbf{0.71}                        & \textbf{9.8}                                                         \\\midrule
SAC for asymptotic feasibility    & \textbf{104.6}$\pm$98.6  & 0.56                        & \textbf{14.7}                                                        \\
SAC for finite-time feasibility& 103.4$\pm$96.4 & \textbf{0.85}                        & 13.9       \\
\bottomrule
\end{tabular}
}
\end{center}
\end{table}

\paragraph{More comparison with different operators.}
In this section, we show the comparison of two neural operators, FNO \citep{li2020fourier} and MNO \citep{li2022learning} for the safety filter performance with $\phi(Y)$ in learning the boundary mapping from control input $U(t)$ to output $Y(t)$ for 1D hyperbolic equation. Note that MNO models have larger model complexity than FNO models.  Different from \Cref{tab:FNO_MNO}, in \Cref{tab:mno_app}, we can see that with weaker BCBF $\phi(Y)$, MNO performs no worse than FNO in feasible rate and reward, showing that larger model complexity will compensate the performance of BCBF in the safety filter framework. 

\begin{table}[t]
\caption{Results of filtering with  BCBF $\phi(Y)$ under different neural operator modeling for first-order transport equation. The boundary feasibility constraint is $Y<1$ for PPO  and $Y<0$ for SAC models.}
\begin{center}
\label{tab:mno_app}
\resizebox{0.9\textwidth}{!}{
    \begin{tabular}{cccc} \toprule
Filtering with different BCBFs                    & \begin{tabular}[c]{@{}c@{}}Reward (mean$\pm$std)\\ (starting at $\sim$-300)\end{tabular}  & \begin{tabular}[c]{@{}c@{}}Feasible Rate\\ (100 episodes)\end{tabular}  &  \begin{tabular}[c]{@{}c@{}}Average Feasible Steps\\ ( 50 control steps)\end{tabular}  \\ \midrule
PPO w. MNO     & \textbf{162.9}$\pm$45.2  & \textbf{0.68 }                       & \textbf{8.7}                                                          \\
PPO w. FNO  & 162.3$\pm$44.5  & 0.63                        & 8.3                                                         \\\midrule
SAC w. MNO      & 103.2$\pm$98.3  & \textbf{0.59}                       & 15.4                                                        \\
SAC w. FNO & \textbf{103.3}$\pm$98.4  & 0.57                       & \textbf{15.7}       \\
\bottomrule
\end{tabular}
}
\end{center}
\end{table}

\begin{figure}[]
    \centering
    \includegraphics[width=0.9\linewidth]{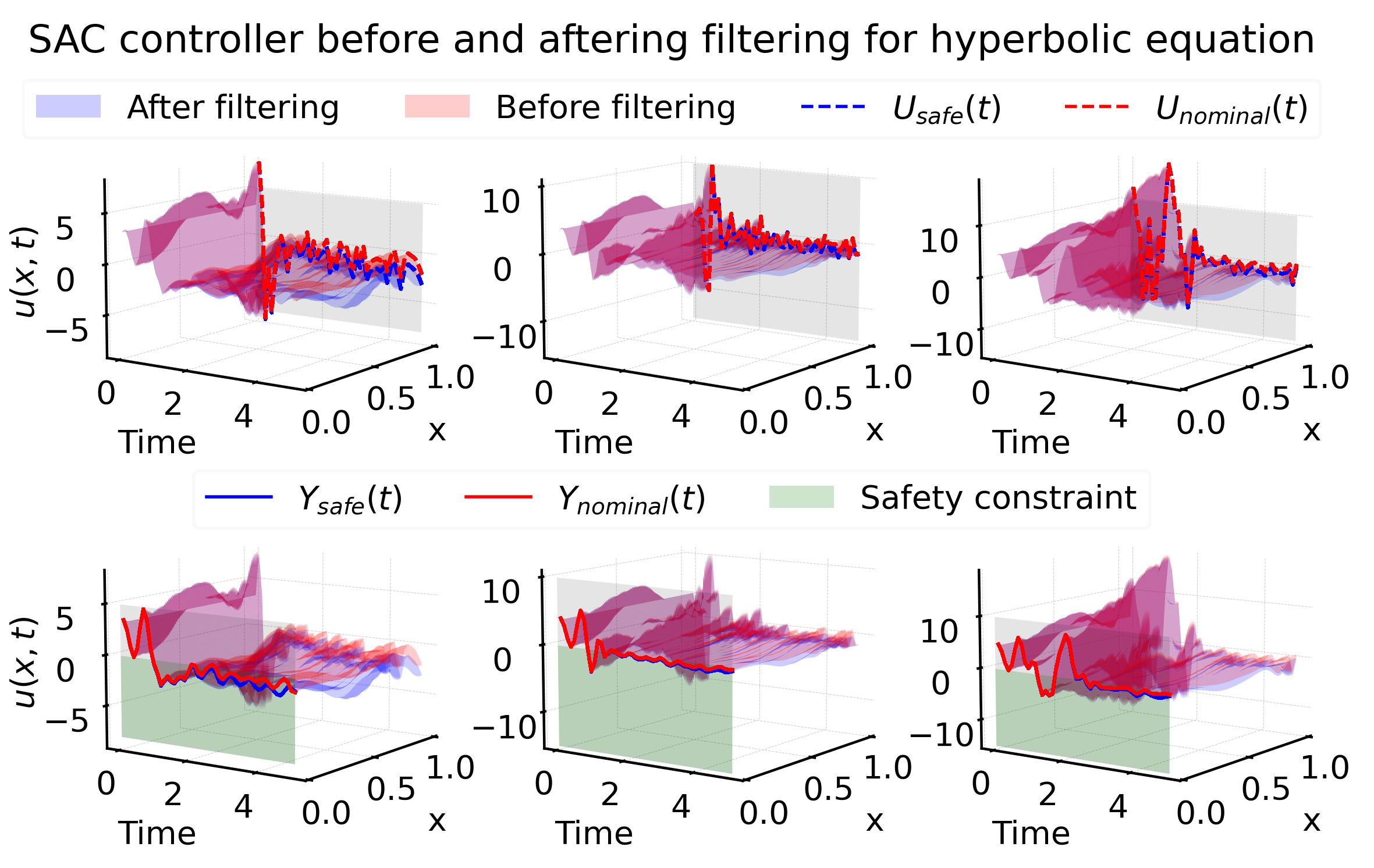}
    \caption{Visualization of state $u(x,t)$ of hyperbolic equation under SAC controller with (in blue) and without (in red) filtering. Boundary control inputs $U(t)$ are in dashed lines and boundary output $Y(t)$ are in solid lines. The boundary constraint $Y(t) < 0$ is in green.}
    \label{fig:visualize_sac}
\end{figure}

\begin{figure}[H]
    \centering
    \includegraphics[width=0.45\linewidth]{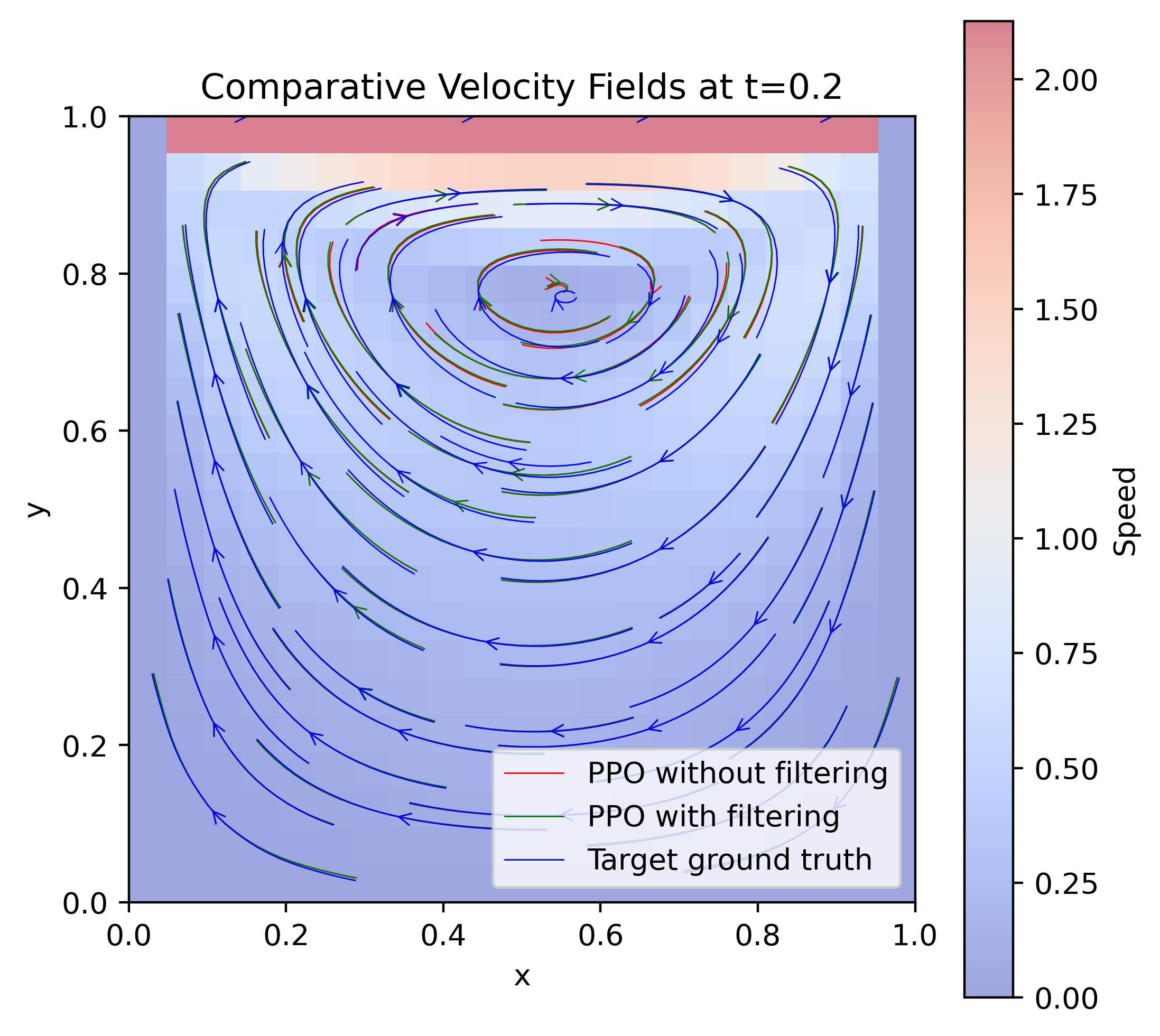}\includegraphics[width=0.45\linewidth]{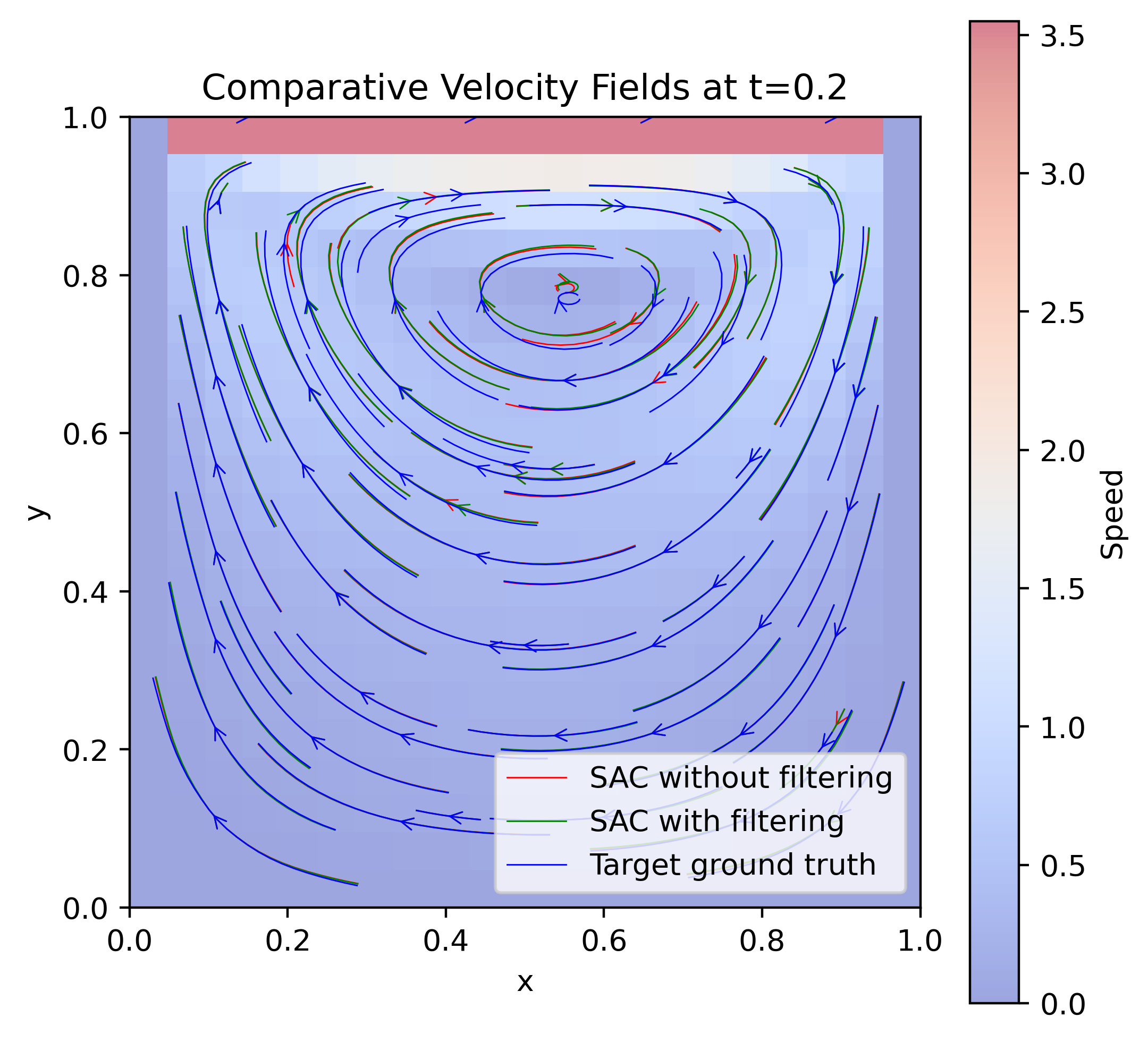}
    \caption{Visualization of tracking performance  with PPO and SAC models before and after filtering with $\phi(t,Y)$ at the end time step of the trajectory for Navier-Stokes equation.}
    \label{fig:visualize_ns}
\end{figure}

\paragraph{More visualization of hyperbolic and Navier-Stokes equations.}
Here, we visualize the trajectories under 1D hyperbolic equation using a SAC controller without and with safety filtering of $\phi(t,Y)$, as shown in \Cref{fig:visualize_sac}. Similar to \ref{fig:visualize_ppo},  for each trajectory, the state value $u(x,t)$ after filtering is lower than that before filtering, i.e., the blue area is lower than the red area. For the output boundary, the filtered one $Y(t)_{\text{safe}}$ in blue solid lines goes towards the constraint $Y(t)<0$ compared to the nominal boundary output $Y(t)_{\text{nominal}}$ in red solid lines, because of the output boundary. The difference is not very large in the last two figures because the threshold is relatively small to keep the stability of the output. As the visualization shows in \Cref{fig:visualize_ns}, it can be seen that the mid-upper high-speed tracking performance is improved compared to the baseline without filtering due to the constraint satisfaction. However, since the output boundary is just one point in the high-speed part, the general performance after filtering is not improved significantly, which is consistent with the findings in \Cref{tab:vanilla}.

\begin{table}[t]
\caption{Comparison of before QP and after QP filtering with different thresholds using $\phi(Y)$ and $\phi(t,Y)$ for PPO model under hyperbolic equation.}
\vspace{-3mm}
\begin{center}
\label{tab:threshold}
\resizebox{0.9\textwidth}{!}{
    \begin{tabular}{cccc} 
    \toprule
    Filtering with $\phi(Y)$ & \begin{tabular}[c]{@{}c@{}}Reward (mean$\pm$std)\end{tabular}  & \begin{tabular}[c]{@{}c@{}}Feasible Rate\end{tabular}  &  \begin{tabular}[c]{@{}c@{}}Average Feasible Steps\end{tabular}  \\ \midrule
Before QP (baseline)                        & 157.90$\pm$37.46  & 0.63  & 7.56   \\ \midrule
After QP with threshold 0.5   & 158.45$\pm$37.82  & 0.65  & 7.49  \\ \midrule
After QP with threshold 2     & 162.26$\pm$44.53  & 0.63  & 8.49  \\ \midrule
After QP with threshold 5     & 114.40$\pm$83.25  & 0.67  & 11.01 \\ \midrule
After QP with threshold 10   & 27.28$\pm$57.62   & 0.57  & 11.30 \\ 
\bottomrule
\end{tabular}
}
\resizebox{0.9\textwidth}{!}{
\begin{tabular}{cccc} \toprule
    Filtering with $\phi(t,Y)$  & \begin{tabular}[c]{@{}c@{}}Reward (mean$\pm$std)\end{tabular}  & \begin{tabular}[c]{@{}c@{}}Feasible Rate\end{tabular}  &  \begin{tabular}[c]{@{}c@{}}Average Feasible Steps \end{tabular}  \\ \midrule
Before QP (baseline)                        & 157.90$\pm$37.46  & 0.63  & 7.56  \\ \midrule
After QP with threshold 0.5   & 158.60$\pm$37.76  & 0.68  & 7.19   \\ \midrule
After QP with threshold 2     & 165.04$\pm$43.73  & 0.71  & 9.80  \\ \midrule
After QP with threshold 5     & 127.18$\pm$82.67  & 0.73  & 12.60  \\ \midrule
After QP with threshold 10    & 28.61$\pm$64.03   & 0.57  & 13.74 \\ 
\bottomrule
\end{tabular}
}
\end{center}
\end{table}

\section{Limitation and Discussion}
Since the proposed method is based on neural operator modeling instead of real PDE dynamics, it does not directly solve the problem of model mismatch which may hurt the safety filtering performance in the implementation. We mark this important point as future work. 
Also, for PDE dynamics with higher-dimensional states, future work is needed to investigate how BCBF can deal with spatially dependent boundaries under complicated boundary constraint settings and safe sets. Another limitation lies in that we do not adopt online safety filtering under the real PDE dynamics, which can be further explored by replacing the offline filtered control input trajectory with the real-time safety filtering at each step in \Cref{alg} in the real-world applications. It is also interesting to omit the iterative filtering by prediction using the one-time filtering for the whole trajectory based on \Cref{eq:pf_G}, which has the challenge of the nonlinear dependence of the neural operator derivative at the initial time. More work can also be explored using neural network verification \citep{wei2024modelverification,yang2024scalable}  to ensure the safety and robustness under input perturbation \citep{cheng2024robust,liu2023towards}.

\end{document}

%% file: math_commands.tex
\usepackage{amsmath,amsfonts,bm}

\def\eqref#1{equation~\ref{#1}}

\def\1{\bm{1}}

\def\vzero{{\bm{0}}}
\def\vone{{\bm{1}}}

\DeclareMathAlphabet{\mathsfit}{\encodingdefault}{\sfdefault}{m}{sl}
\SetMathAlphabet{\mathsfit}{bold}{\encodingdefault}{\sfdefault}{bx}{n}

\def\gD{{\mathcal{D}}}

\def\gG{{\mathcal{G}}}

\def\gI{{\mathcal{I}}}
\def\gJ{{\mathcal{J}}}

\def\gL{{\mathcal{L}}}

\def\gP{{\mathcal{P}}}
\def\gQ{{\mathcal{Q}}}

\def\gS{{\mathcal{S}}}
\def\gT{{\mathcal{T}}}

\def\gX{{\mathcal{X}}}

\def\sR{{\mathbb{R}}}

\DeclareMathOperator*{\argmin}{arg\,min}

%% file: intro.tex
\section{Introduction}
Partial differential equations (PDEs) characterize the most fundamental laws of the continuous dynamical systems in the physical world \citep{evans2022partial,perko2013differential}. Non-analytical PDE dynamics are often involved in complicated science and engineering problems of computational fluid dynamics \citep{kochkov2021machine}, computational mechanics \citep{samaniego2020energy},  robotics \citep{heiden2021neuralsim}, etc. Recently, neural networks have  largely  boosted the study of numerical PDE solvers using data-driven methods, simulating and characterizing the dynamics \citep{raissi2019physics, brunton2024promising,kovachki2023neural}. However, the PDE control problem remains challenging without any prior knowledge about underlying PDE equations, serving as a huge gap from understanding science to solving engineering problems \citep{yu2024learning}.

Recent pioneer works \citep{bhan2024pde,zhang2024controlgym} provide various formulations of PDE control problems and multiple benchmark settings, either in-domain control \citep{zhang2024policy} or boundary control \citep{bhan2023neural}. Since it is easier to control the PDE boundary in the real world, following \cite{bhan2024pde}, we focus on the PDE boundary control setting where the control signal essentially serves as the boundary condition and the unknown PDE dynamics itself remains unchanged. Model-based PDE boundary control has been studied for years, and backstepping-based methods have been applied to different PDE dynamics \citep{krstic2008boundary}. Nevertheless, the model-based methods cannot work well under the unknown PDE dynamics, suffering from significant model mismatch. Model-free reinforcement learning (RL) controllers \citep{schulman2017proximal,haarnoja2018soft} have shown impressive results in the benchmark \citep{bhan2024pde} compared to the model-based control methods \citep{pyta2015optimal}. 

\begin{figure}
    \centering
    \includegraphics[width=0.8\linewidth]{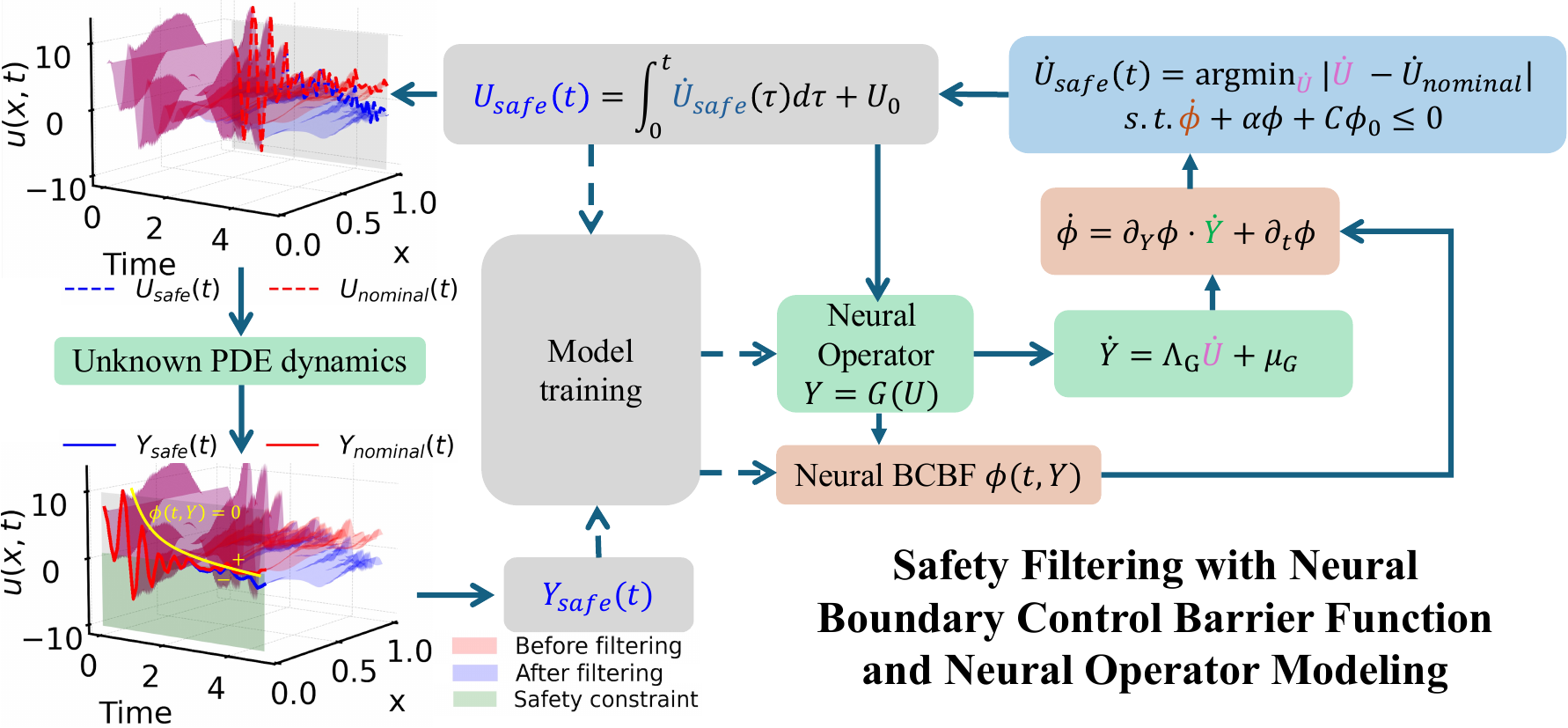}
    \vspace{-3mm}
    \caption{Overview of our safety filtering method for PDE boundary control with neural BCBF. Solid line arrows denote the safety filtering, while dashed ones denote the model training.}
    \label{fig:overview}
    \vspace{-8mm}
\end{figure}

Besides, constraint satisfaction is of great importance for the PDE boundary control problems, but current safe PDE control methods are typically backstepping-based  and require knowledge about the PDE dynamics \citep{krstic2006nonovershooting,li2020mean,koga2023safe,wang2023safe}. The constraint considered in this paper is called \textit{boundary feasibility}, which characterizes whether the boundary output falls into and stays within the safe set at the end of the finite-time trajectory, and can be understood as the constraint of finite-time convergence.
Under ordinary differential equations (ODEs) setting, neural network parameterized control Lyapunov/barrier functions (CLF/CBFs) have been adopted to ensure the
 convergence and safety of learning-based controllers \citep{boffi2021learning,dawson2023safe,chang2019neural,mazouz2022safety}, based on the Markov property of the dynamics at each step
 , i.e., the change of state only depends on the current state and control input. 
 However,  the Markov assumption does not generally hold for PDE boundary control due to infinite-dimensional unobserved states along the spatial axis. It is also challenging to bypass the unknown PDE dynamics to 
 to find the boundary control input at each step for trajectory-wise convergence over boundary output constraint.

To this end, we introduce a new framework to achieve \textit{boundary feasibility} within a given safe set for the PDE boundary control problem, as shown in \Cref{fig:overview}. More specifically, we propose neural boundary control barrier functions (BCBFs) over the boundary output to enable the incorporation of the time variable with a finite-time convergence guarantee.
Then, we adopt a neural operator to directly learn the mapping from boundary input to output as a transfer function. Combining well-trained neural BCBF and neural operator, we show a linear dependence between boundary feasibility condition and the derivative of boundary control input, making the safety filtering possible by projecting the actions from the nominal RL controller to the safe boundary control input set using quadratic programming (QP). 
We conduct experiments on multiple PDE benchmarks and show our plug-and-play filtering superiority over vanilla and constrained RL controllers regarding general performance and constraint satisfaction. To the best of our knowledge, we are the first to study safe boundary control with unknown PDE dynamics. More related work is discussed in 
\Cref{sec:related}.
We summarize our contributions below.
\begin{itemize}[leftmargin=0.4cm]
    \item We propose a new PDE safe  control framework with a neural boundary control barrier function to guarantee the boundary feasibility of the boundary output within a given safe set.
    \item We model the control input and output mapping through a neural operator as a transfer function and prove that it can be used for safety filtering by solving quadratic programming.
    \item We show that the add-on performance after safety filtering is better than both vanilla and constrained RL controllers in boundary feasibility rate and time steps on multiple PDE environments.
\end{itemize}

%% file: formulation.tex
\section{Problem Formulation}
Following the PDE boundary control setting \citep{bhan2024pde}, 
we consider the state $u(x,t):\gX\times\gT\rightarrow\gS\subset \sR$ from the continuous function space $C(\gX\times\gT;\sR)$ governed by underlying closed-loop partial differential equation (PDE) dynamics defined on normalized $n$-dimensional spatial domain $\gX=[\vzero,\vone]:=[0,1]^n\subset\sR^n$ and temporal domain $\gT=[0,T]\subset \sR^+$ as follows,
\begin{align}
    \frac{\partial u}{\partial t} = \gD(u, \frac{\partial u}{\partial x},\frac{\partial^2 u}{\partial x^2},\dots, U(t)), x\in\gX, t\in\gT, u\in\gS,  \label{eq:pde_dynamics}
\end{align}
where $\gD$ is the PDE system dynamics and $U(t)$ is the  control signal as the boundary condition. Without loss of generality, we focus on the Dirichlet boundary control input as $U(t):=u(\vone, t)$ with constant initial condition $u(x,0)\equiv U(0)\in \gS$. Instead  of optimizing boundary input $U(t)$ to track or stabilize full-state observation trajectory $u(x,t)$ \citep{bhan2024pde}, we aim to find $U(t)$ that guarantees the \textit{boundary feasibility} of boundary output $Y(t):=u(\vzero,t)$ within the given user-specified safe set $\gS_0\subset \gS$ over $\gT$, i.e., $\exists t_0\in\gT, \forall t\geq t_0, Y(t)\in \gS_0$. 
Note that the boundary states can be generalized to any spatially marginalized state-related trajectories. 
More formally, we define \textit{boundary feasibility} as follows in PDE dynamics.
\begin{definition}[Boundary Feasibility for Finite-time Constraint Satisfaction]
\label{def:pf}
    With state $u(x,t)$ subjected to closed-loop PDE dynamics in \Cref{eq:pde_dynamics} with the boundary control input $U(t)$,  the boundary control output $Y(t)$ is defined to be feasible over $\gT$ within the given user-specified safe set $\gS_0\in\gS$ if the following holds,
    \begin{align}
    \label{eq:pf}
        \exists t_0\in\gT, \forall  t_0 \leq t\leq T, Y(t):=u(\vzero,t)\in \gS_0, \text{ where } u(\vone,t)=U(t), u(x,0)\equiv U(0).
    \end{align}
\end{definition}
With boundary input and output trajectory pairs $\{[U_k(t),Y_k(t)], k=1,2,\dots, K\}$ from the unknown PDE dynamics,we formulate the problem for this paper as follows.
\begin{prbl}
\label{prob:prob}
    Given $K$ collected boundary input and output  trajectory pairs $\{[U_{k,m},Y_{k,m}], k=1,2,\dots, K,m=1,2,\dots,M\}$ with $M$-point temporal discretization, under consistent initial condition $u_k(x,0)\equiv U_k(0)$ from unknown but time-invariant PDE dynamics in \Cref{eq:pde_dynamics}, we aim to find boundary control input $U(x)$ that guarantees boundary feasibility of boundary output $Y(t)$  with user-specified safe set $\gS_0$ in Definition \ref{def:pf}.
\end{prbl}

%% file: method.tex
\section{Methodology}
\subsection{Neural Barrier Function for PDE Boundary Control}
Boundary feasibility aims to find control input $U(t)$ for the constraint satisfaction of the marginalized output boundary  $Y(t):=u(\vzero,t)$ from the underlying PDE dynamics with spatially-continuous unobservable state $u(x,t)$, which is challenging for conventional state-dependent-only CBFs.
Hence,
inspired by \cite{garg2021robust}, 
we propose the neural boundary control barrier function (neural BCBF), explicitly incorporating time $t$ into neural network parameterized function $\phi(t, Y):\gT\times\gS\rightarrow\sR$ for the time-dependent zero-sublevel set $\gS_{\phi,t}:=\{Y(t)\mid \phi(t,Y(t))\leq 0\}$. Note that the conventional CBF $\phi(Y)$ can be viewed as a specially case of  BCBF $\phi(t, Y)$ where $t$ remains constant. 
Another challenge is that the boundary feasibility in \Cref{eq:pf} for PDE boundary control is defined on finite time domain $\gT=[0,T]$, which requires a higher convergence rate to the safe set than the original asymptotic CBF \citep{ames2014control} like fixed-time stability in \cite{polyakov2011nonlinear, garg2021characterization}.  The following theorem shows the feasibility of boundary control output $Y(t)$ within the user-specified safe set $\gS_0$ under control signal $U(t)$.
\begin{theorem}[Boundary Feasibility with Boundary Control Barrier Function]
\label{thm:pf_thm}
For the state $u(x,t)$ from the closed-loop PDE dynamics with boundary control input $U(t)=u(\vone,t), u(x,0)\equiv U_0$, the boundary feasibility of boundary  output $Y(t)=u(\vzero,t)$ over $\gT=[0,T]$ within user-specified safe set $\gS_0$ is guaranteed with neural BCBF $\phi(t,Y)$ if the following holds $\forall t\in\gT$
\begin{align}
\label{eq:cbf_pf}
    \left(\gS_{\phi,t}:=\{Y\mid \phi(t,Y)\leq 0\}\subseteq \gS_0\right)\bigwedge \left(\partial_Y\phi\cdot \frac{d Y}{d t} +\partial_t\phi+ \alpha \phi(t,Y) + C_{\alpha,T}\phi(0,U_0) \leq 0\right),
\end{align}
where $C_{\alpha,T} := \frac{\alpha}{e^{\alpha T}-1}>0$ is a constant for  finite-time convergence. 
\end{theorem}
\begin{proof}
    With the sublevel set  $\gS_{\phi,t}$ being the subset of $\gS_0$, i.e., $\gS_{\phi,t}:=\{Y\mid \phi(t,Y)\leq 0\}\subseteq \gS_0$, it is sufficient to prove $\exists t_0 \in [0,T], s.t. \forall t\in[t_0, T], \phi(t,Y(t)) \leq 0$.
    Now denote $\psi(t):= \phi(t,Y(t))$, by initial constant boundary condition $Y(0)=U_0$, the following equivalent inequalities hold,
        $$\partial_Y\phi\cdot \frac{d Y}{d t} +\partial_t\phi+ \alpha \phi(t,Y) + C_{\alpha,T}\phi(0,Y(0)) \leq 0
        \Longleftrightarrow \frac{d(e^{\alpha t}\psi(t)+\frac{C_{\alpha,T}\psi(0)}{\alpha}e^{\alpha t})}{dt} \leq 0 $$
     So the function $e^{\alpha t}\psi(t)+\frac{C_{\alpha,T}\psi(0)}{\alpha}e^{\alpha t}$ is non-increasing over $t\in[0,T]$. By $T>0$, we have
     \begin{align*}
         [e^{\alpha t}\psi(t)+\frac{C_{\alpha,T}\psi(0)}{\alpha}e^{\alpha t}]|_{t=T} < [e^{\alpha t}\psi(t)+\frac{C_{\alpha,T}\psi(0)}{\alpha}e^{\alpha t}]|_{t=0}
\Longleftrightarrow \psi(T)  = \phi(T, Y(T))  < 0 
     \end{align*}
     So at least at $t_0=T$, $\phi(t_0, Y(t_0)) < 0$, which concludes the proof of boundary feasibility of boundary  output $Y(t)=u(\vzero,t)$ over $\gT=[0,T]$ in  Definition \ref{def:pf}.
\end{proof}
The full proof can be found in 
\Cref{sec:proof_pf}. 
Note that if $\phi(0, U_0)\leq 0$, the forward invariance \citep{ames2019control} can be obtained via $T\rightarrow \infty$. With the $M$-point temporal discretization of collected boundary input and output trajectory $\{[U_{k,m},Y_{k,m}],k=1,\dots,K, m=1,\dots,M\}$, $\gS_{\phi,t}\subseteq\gS_0$ in  \Cref{eq:cbf_pf} induces the loss below following  \cite{dawson2022safe} 
\begin{align}
\label{eq:loss_S}
    \gL_\gS = \sum_{k=1}^K\sum_{Y_{k,m}\in\gS_0}[\phi(t_m,Y_{k,m})]_+ +\sum_{k=1}^K\sum_{Y_{k,m}\notin\gS_0}[-\phi(t_m,Y_{k,m})]_+ , \text{ with } [\cdot]_+:=\max\{0,\cdot\}.
\end{align}
 However, it is challenging to find $\nicefrac{dY(t)}{dt}$ involved in \Cref{eq:cbf_pf} over the discrete time samples since the boundary output $Y(t)=u(\vzero,t)$ is governed by the unknown closed-loop PDE dynamics with the boundary condition $U(t)=u(\vone,t)$. Besides, it is also non-trivial to find the boundary  feasibility condition over boundary control input $U(t)$ for safety filtering due to non-Markov property. Therefore, we adopt the neural operator to learn the boundary input-output mapping as a neural transfer function to further mitigate the non-Markov issue in PDE boundary control problems with unknown PDE dynamics.

\subsection{Learning Neural Operator for Input-output Boundary Mapping}
\label{sec:NO_formulate}
Different from current applications of neural operators in learning PDE solutions by temporal mapping \citep{li2020fourier,li2020neural,li2022learning}, we propose to adopt neural operator $\gG_\theta:\{U: \gT\rightarrow\gS\}\mapsto\{Y: \gT\rightarrow\gS\}$ to model the spatial boundary mapping from input to output of the unknown closed-loop PDE dynamics in \Cref{eq:pde_dynamics}, i.e., $Y(t)=u(\vone,t)=\gG_\theta(U)(t)=\gG_\theta(u(\vzero,t))(t)$. Following \cite{kovachki2023neural} under the setting of same  Lebesgue-measurable domain $\gT$ for hidden layers,  the neural operator is defined as $\gG_\theta=\gQ\circ\gI_{L-1}\circ\dots\circ\gI_0\circ\gP$, including pointwise lifting mapping $\gP:\{U: \gT\rightarrow\gS\}\mapsto\{v_0: \gT\rightarrow\sR^{d_{v_0}}\}$, iterative kernel integration layers $\gI_l: \{v_l: \gT\rightarrow\sR^{d_{v_l}}\}\mapsto\{v_{l+1}: \gT\rightarrow\sR^{d_{v_{l+1}}}\}, l=0,\dots,L-1$, and the pointwise projection mapping $\gQ:\{v_L: \gT\rightarrow\sR^{d_{v_L}}\}\mapsto\{Y: \gT\rightarrow\gS\}$. Specifically, the $l$-th kernel integration layer follows the following form with commonly-used integral kernel operator \citep{li2020fourier,li2020neural,li2022learning},
\begin{align}
\label{eq:int_layer}
    v_{l+1}(t) = \gI_l(v_l)(t) = \sigma_{l+1}\left(W_lv_l(t)+\int_\gT\kappa^{(l)}(t,s)v_l(s)ds+b_l(t)\right), l=0,1,\dots,L-1,
\end{align}
where $\sigma_{l+1}:\sR^{d_{v_{l+1}}}\rightarrow\sR^{d_{v_{l+1}}}$ is the activation function, $W_l\in\sR^{d_{v_{l+1}}\times d_{v_l}}$ is the local linear operator, $\kappa^{(l)}\in C(\gT\times\gT;\sR^{d_{v_{l+1}}\times d_{v_l}})$ is the  kernel function for integration, and $b_l\in C(\gT;\sR^{d_{v_{l+1}}})$ is the bias function. 
Besides, since lifting and projection operators $\gP,\gQ$ are pointwise local maps as special Nemitskiy operators \citep{dudley2011concrete,kovachki2023neural}, i.e. there exist equivalent functions $ P: \gS\rightarrow\sR^{d_{v_0}}, Q: \sR^{d_{v_L}}\rightarrow\gS$ such that $\gP(U)(t)=P(U(t)), \gQ(v_L)(t)=Q(v_L(t)),\forall t \in \gT$. Therefore, combining \Cref{eq:int_layer}, we  explicitly show the boundary mapping from  control input $U(t)$ to  output $Y(t)$ below, making them possible to be directly connected as $Y(t)=\gG_\theta(U)(t)$,
\begin{align}
\label{eq:no_forward}
    Y(t)=\gG_\theta(U)(t)=Q(v_L(t)), v_{l+1}(t) = \gI_l(v_l)(t) \text{ in \Cref{eq:int_layer}}, v_0(t) =P(U(t)),
\end{align}
where $P,Q,W_l,\kappa^{(l)},b_l,l=0,1,\dots,L-1$  parameterized with neural networks $\theta$ and compose the neural operator $Y(t)=G_\theta(U)(t)$. Given boundary input and output $M$-step temporally discretized $K$ trajectory pairs $\{[U_{k,m},Y_{k,m}], k=1,2,\dots, K,m=1,2,\dots, M\}$, $G_\theta$ and neural BCBF $\phi$ can be optimized together based on empirical-risk minimization using the following loss function,
\begin{align}
\label{eq:total_loss}
    &\min_{\theta,\phi}\lambda_\gG\gL_\gG + \lambda_\gS\gL_\gS + \lambda_{BF}\gL_{BF}, \text{ where } \gL_\gG=\sum_{k=1}^K\sum_{m=1}^M\|Y_{k,m} - \gG_\theta(U_k)(t_m)\|^2, \gL_\gS \text{ in \cref{eq:loss_S}}, \notag \\ & \gL_{BF}=\sum_{k=1}^K\sum_{m=1}^M[\partial_{Y_{k,m}}\phi\cdot \frac{d \gG_\theta(U_k)(t)}{d t}\mid_{t=t_m} +\partial_{t_m}\phi+ \alpha \phi(t_m,Y_{k,m}) + C_{\alpha,T}\phi(0,U_{k,0})]_+, 
\end{align}
and  $[\cdot]_+:=\max\{0,\cdot\},, \lambda_\gG, \lambda_\gS,\lambda_{BF}$ are weight hyperparameters for $\gL_\gG,\gL_\gS,\gL_{BF}$, respectively. The loss for neural operator learning $\gL_\gG$ is based on  \Cref{eq:no_forward}, and the boundary feasibility (BF) loss of $\gL_{BF}$ is based on \Cref{eq:cbf_pf} with the replacement of $\nicefrac{dY(t)}{dt}$ with $\nicefrac{d\gG_\theta(U)(t)}{dt}$, which will be detailed in the next section.

\subsection{Safety Filtering with Quadratic Programming}
Once the boundary input-output mapping is modeled by neural operator $\gG_\theta$, the boundary output $Y(t)$ is directly related to boundary input $U(t)$ from trajectory to trajectory, bypassing the non-Markov property and the unknown closed-loop dynamics in \Cref{eq:pde_dynamics}.
We first find the derivative of boundary output $Y(t)$ w.r.t $t$ based on neural operator $Y(t) = \gG_\theta(U)(t)$. Applying chain rule to \Cref{eq:no_forward}, the following derivatives hold,
\begin{align}
    \label{eq:no_backward}
    \frac{dY(t)}{dt} = \nabla Q^\top\frac{dv_L(t)}{dt}, \frac{dv_{l+1}(t)}{dt} = \gJ_l(\frac{dv_l}{dt})(t), l = L-1, \dots,0, \frac{v_0(t)}{dt} = \nabla P^\top \frac{dU(t)}{dt},
\end{align}
where  the  derivative of kernel integration layer $\gJ_l: \{\frac{v_l}{dt}: \gT\rightarrow\sR^{d_{v_l}}\}\mapsto\{\frac{v_{l+1}}{dt}: \gT\rightarrow\sR^{d_{v_{l+1}}}\},l=0,1,\dots,L-1$ can be found through the derivative  of \Cref{eq:int_layer} in a recursive form below,
\begin{align}
    \label{eq:de_int_layer}
    \frac{dv_{l+1}(t)}{dt} = \gJ_l(\frac{dv_l}{dt})(t) = \text{Diag}(\sigma_{l+1}')\left(W_l\frac{dv_l(t)}{dt}+\int_\gT\frac{\partial \kappa^{(l)}(t,s)}{\partial t}v_l(s)ds+\frac{db_l(t)}{dt}\right).
\end{align}
By combining \Cref{eq:no_backward} and \Cref{eq:de_int_layer}, we have the following theorem to show how the boundary control input $U(t)$ can be chosen to guarantee the boundary feasibility of boundary output $Y(t)$ modeled by neural operator $\gG_\theta$.

\begin{theorem}[Boundary Feasibility with Neural Operator]
\label{thm:certificate}
Assuming the neural operator $\gG_\theta$ as an exact map from boundary input $U(t)$ to output $Y(t)$
for  an unknown closed-loop PDE dynamics  without model mismatch, the boundary control input $U(t)$  is guaranteed to induce  boundary feasibility of output  $Y(t)$ over $\gT=[0,T]$ within the sublevel set of neural BCBF $\phi$  if $ U(t)$ satisfies 
\begin{align}
\label{eq:pf_G}
    \partial_Y\phi(t,\gG_\theta(U)) \frac{d \gG_\theta(U)(t)}{d t} +\partial_t\phi(t,\gG_\theta(U))+ \alpha \phi(t,\gG_\theta(U)) + C_{\alpha,T}\phi(0, U(0)) \leq 0, \forall t\in\gT
\end{align}
 where $C_{\alpha,T} = \frac{\alpha}{e^{\alpha T}-1}$, and $\frac{d \gG_\theta(U)(t)}{d t}$ can be  found below with $\prod_1^0 (\cdot):=1$,
\begin{align}
\label{eq:de_no_full}
&\frac{d \gG_\theta(U)(t)}{d t} = \nabla Q^\top\prod_{l=0}^{L-1}\left(\text{Diag}(\sigma_{L-l}')W_{L-1-l}\right)\nabla P^\top \frac{dU(t)}{dt} + \nabla Q^\top\text{Diag}(\sigma_L') \sum_{i=0}^{L-1} \left(\big[\prod_{j=1}^i W_{L-j}\right. \notag\\&\left. \text{Diag}(\sigma_{L-j}')\big] \left(\int_\gT\frac{\partial \kappa^{(L-1-i)}(t,s)}{\partial t}v_{L-1-i}(s)ds+\frac{db_{L-1-i}(t)}{dt}\right)\right) = \Lambda_\theta(t) \dot U(t) + \mu_\theta(t).
\end{align}
\end{theorem}
The proof of \Cref{eq:pf_G} is based on \Cref{thm:certificate} and \Cref{eq:de_no_full} can be derived by recursively applying \Cref{eq:de_int_layer} to \Cref{eq:no_backward}. Please check
\Cref{sec:proof_cert}
for full proof.
\begin{remark}
    We remark that if the sublevel set of neural BCBF $\phi$ is a subset of user-specified safe set $\gS_0$,  and there is no model mismatch between neural operator $Y(t)=\gG_\theta(U)(t)$ and unknown closed-loop PDE dynamics,  Theorem \ref{thm:certificate} is equivalent to Theorem \ref{thm:pf_thm}. Then the boundary control input $U(t)$ satisfying \Cref{eq:pf_G} is guaranteed to induce the boundary feasibility of boundary output $Y(t)$ within the user-specified safe set $\gS_0$. 
\end{remark}
Based on the affine property of $\dot U(t)$ in \Cref{eq:de_no_full}, we formulate the following quadratic programming  with neural BCBF $\phi$ and neural operator $\gG_\theta$ as a safety filter for $\dot U_{\text{nominal}}(t),\forall t \in \gT$,
\begin{align}
    \label{eq:safety_filter}
     &\dot U_{\text{safe}}(t) = \argmin_{\dot U \in\sR}\|\dot U - \dot U_{\text{nominal}}(t)\| \\
    &s.t.~ \partial_Y\phi(t,Y) \left(\Lambda_\theta(t) \dot U + \mu_\theta(t)\right) +\partial_t\phi(t,Y)+ \alpha \phi(t,Y) + C_{\alpha,T}\phi(0,U_{\text{nominal}}(0)) \leq 0,
\end{align}
where $C_{\alpha,T} = \frac{\alpha}{e^{\alpha T}-1}$ and $ \Lambda_\theta(t),\mu_\theta(t)$ can be found in \Cref{eq:de_no_full}. 
Based on $\dot U_{\text{safe}}(t)$ at each step $t$, we update the potential boundary control input  $U_{\text{safe}}(t)$ as $U_{\text{safe}}(t) = \int_{0}^t \dot U_{\text{safe}}(\tau) d\tau + U_{\text{nominal}}(0)$, so that the predicted boundary output $Y_{\text{predict}}(t)=\gG_\theta(U_{\text{safe}})(t)$ can be found by the neural operator $\gG_\theta$. Therefore, the next QP update can be solved for $\dot U_{\text{safe}}$ at the next time by \Cref{eq:safety_filter}. Note that we let $\dot U_{\text{safe}}=\dot U_{\text{nominal}}$ for the unfiltered time steps during the QP iteration. The discrete-time implementation of the safety filtering procedure is shown in \Cref{alg}. 
To accommodate the advection-dominated problems like the 1D hyperbolic problem or Navier Stokes, where the propagation speed from input to output boundary is not infinite, we predict the whole input and potentially delayed output trajectory through the neural operator at each step during the safety filtering.
We adopt the predicted $Y(t)$ from the neural operator after each filtering step, 
and the filtering threshold is detailed as a workaround for the model mismatch in 
\Cref{sec:exp_app}, 
along with a discussion on how approximation errors affect safety filtering.
\begin{figure}[t]
\vspace{-4mm}
\begin{algorithm}[H]
\caption{Safety Filtering Procedure for Discrete-time Implementation}
\begin{algorithmic}[1]
\label{alg}
\STATE \textbf{Input:} Initial and nominal control input  $U^{\text{nominal}}_{0:M}$, neural operator $\gG$, neural BCBF $\phi$
\STATE \textbf{Output:} Filtered safe control input  $U^{\text{safe}}_{1:M}$
\STATE Initialize $\Delta U^{\text{safe}}_{1:M} = \Delta U^{\text{nominal}}_{1:M}\gets U^{\text{nominal}}_{1:M}-U^{\text{nominal}}_{0:M-1}, Y^{\text{predict}}_{1:M} \gets \gG(U^{\text{nominal}}_{1:M})$ 
\FOR{$m = 1:M$ } 
    \STATE Find $\Delta U^{\text{safe}}_m$ through QP in \Cref{eq:safety_filter} based on $\Delta U^{\text{nominal}}_m,Y^{\text{predict}}_{1:M}, \gG, \phi,U^{\text{nominal}}_0$
    \STATE Update $U^{\text{safe}}_{1:M} \gets \sum_{i=1}^m \Delta U^{\text{safe}}_i + U^{\text{nominal}}_{0}$ 
    \STATE Update $Y^{\text{predict}}_{1:M}\gets\gG(U^{\text{safe}}_{1:M})$
\ENDFOR
\STATE \textbf{return} $U^{\text{safe}}_{1:M}$
\end{algorithmic}
\end{algorithm}
\vspace{-10mm}
\end{figure}




%% file: experiment.tex
\section{Experiment}
In this section, we aim to answer the following two questions: How does the proposed plug-and-play safety filtering perform based on the vanilla and constrained RL controllers in unknown PDE dynamics? How do different types of barrier functions, convergence criteria, and neural operator modeling influence the performance of the proposed safety filtering? We answer the first question in \Cref{sec:result_compare} and the second one in \Cref{sec:ablation}, following the  setup of model training and evaluation metrics. 
\Cref{sec:exp_app}
gives more details and results.
\subsection{Experimental Setup}
\begin{table}[t]
\caption{Comparison of vanilla models w/o and w/ safety filtering under multiple environments. }
\vspace{-7mm}
\begin{center}
\label{tab:vanilla}
\resizebox{0.8\textwidth}{!}{
    \begin{tabular}{cccc} \toprule
1D hyperbolic  equation                   & \begin{tabular}[c]{@{}c@{}}Reward (mean$\pm$std)\\ (starting at $\sim$-300)\end{tabular}  & \begin{tabular}[c]{@{}c@{}}Feasible Rate\\ (100 episodes)\end{tabular}  &  \begin{tabular}[c]{@{}c@{}}Average Feasible Steps\\ ( 50 control steps)\end{tabular}  \\ \midrule
 PPO in \cite{bhan2024pde}     & 157.9$\pm$37.5  & 0.63                        & 7.6                                                          \\
 PPO with filtering & {165.0}$\pm$43.7  & \textbf{0.71 }                       & \textbf{9.8}                                                          \\\midrule
SAC in \cite{bhan2024pde}      & {106.2}$\pm$98.7  & 0.78                        & 12.4                                                         \\
SAC with filtering & 103.4$\pm$96.4 & \textbf{0.85}                        & \textbf{13.9}       \\
 \toprule
1D parabolic equation                        & \begin{tabular}[c]{@{}c@{}}Reward (mean$\pm$std)\\ (starting at $\sim$0)\end{tabular}  & \begin{tabular}[c]{@{}c@{}}Feasible Rate\\ (100 episodes)\end{tabular}  &  \begin{tabular}[c]{@{}c@{}}Average Feasible Steps\\ ( 1000 control steps)\end{tabular} \\ \midrule
 PPO in \cite{bhan2024pde}      & 164.5$\pm$20.7  & 0.60                        & 155.0                                                         \\
 PPO with filtering  & {168.2}$\pm$23.5  & \textbf{0.81}                        & \textbf{507.0}                                                        \\\midrule
SAC in \cite{bhan2024pde}       & 156.5$\pm$6.2  & 0.72                        &     118.4                                                     \\
SAC with filtering  & {157.5}$\pm$6.8   & \textbf{0.87}                        &  \textbf{449.8}     \\
\toprule
2D Navier-Stokes equation                       & \begin{tabular}[c]{@{}c@{}}Reward (mean$\pm$std)\\ (starting at $\sim$-100)\end{tabular}  & \begin{tabular}[c]{@{}c@{}}Feasible Rate\\ (100 episodes)\end{tabular}  &  \begin{tabular}[c]{@{}c@{}}Average Feasible Steps\\ ( 200 control steps)\end{tabular} \\ \midrule
 PPO in \cite{bhan2024pde}     & {-5.37}$\pm$0.01  & 0.86                        & 2.0                                                         \\
 PPO with filtering & -5.72$\pm$0.17  & \textbf{0.99}                       & \textbf{32.0}                                                          \\\midrule
SAC in \cite{bhan2024pde}       & {-18.05}$\pm$1.13  & 0.80                        & 17.5                                                       \\
SAC with filtering & -18.36$\pm$1.25  & \textbf{0.85}                        & \textbf{21.3}       \\
\bottomrule
\end{tabular}
}
\end{center}
\vspace{-7mm}
\end{table}

\begin{table}[]
\caption{Comparison of constrained RL models w/o and w/ safety filtering for 1D hyperbolic PDE. }
\vspace{-3mm}
    \centering
    \label{tab:safeRL}
    \resizebox{0.95\textwidth}{!}{
    \begin{tabular}{cccccc}
    \toprule
Constrained RL Models                         & \begin{tabular}[c]{@{}c@{}}Reward (mean$\pm$std)\\ (starting at $\sim$-300)\end{tabular}     & \multicolumn{2}{c}{\begin{tabular}[c]{@{}c@{}}Feasible Rate under $Y$ constraints\\ (100 episodes)\end{tabular}} & \multicolumn{2}{c}{\begin{tabular}[c]{@{}c@{}}Average Feasible Steps\\ ( 50 control steps)\end{tabular} } \\ \midrule
CPO  \citep{achiam2017constrained}              & 168.7$\pm$28.8 & 0.88  (\textit{Y}\textless{}1)  & 0.52(\textit{Y}\textless{}0)     & 11.2 (\textit{Y}\textless{}1)         & 4.2 (\textit{Y}\textless{}0)            \\
CPO with filtering & {168.8}$\pm$28.6 & \textbf{0.89}  (\textit{Y}\textless{}1)  & \textbf{0.56} (\textit{Y}\textless{}0)     & \textbf{14.8} (\textit{Y}\textless{}1)         & \textbf{4.7} (\textit{Y}\textless{}0)            \\ \midrule
SAC-Lag \citep{ha2020learning}               & {110.9}$\pm$92.1 & 0.84 (\textit{Y}\textless{}0)   & 0.50 (\textit{Y}\textless{}-0.5)  & \textbf{20.8} (\textit{Y}\textless{}0)         & \textbf{3.1} (\textit{Y}\textless{}-0.5)         \\
SAC-Lag with filtering & 107.6$\pm$90.3 & \textbf{0.90} (\textit{Y}\textless{}0)   & \textbf{0.67} (\textit{Y}\textless{}-0.5)  & 18.9 (\textit{Y}\textless{}0)         & 2.9 (\textit{Y}\textless{}-0.5)  \\\bottomrule      
\end{tabular} 
}
\vspace{-6mm}
\end{table}

\paragraph{Environments and model-free controllers.} We adopt the challenging PDE boundary control environments and the model-free reinforcement learning (RL) models from \cite{bhan2024pde} to conduct our experiment. More specifically, the three environments include the unstable 1D hyperbolic (transport) equation, 1D parabolic  (reaction-diffusion) equation and 2D nonlinear Navier-Stokes equation, where the last one is for tracking task and others are for stabilization task.  Since our setting in Problem \ref{prob:prob} does not have prior to the PDE equations, we choose the vanilla PPO \citep{schulman2017proximal} and SAC \citep{haarnoja2018soft}, and constrained RL models CPO  \citep{achiam2017constrained} and SAC-Lag \citep{ha2020learning}  as the baselines in each environment for fair comparisons. The boundary control inputs are consistent with \cite{bhan2024pde}. For 1D environments, the boundary output for the hyperbolic PDE is $Y(t) = u(0,t)$ and the  boundary output for the parabolic PDE $Y(t) = u(0.5,t)$. For the 2D environment, the boundary output is $Y(t)=u(0.5, 0.95, t)$, which has the maximum speed over 2D plane.
The boundary feasibility constraints are detailed in 
\Cref{sec:exp_setup_app}.
With the PDE controllers in \cite{bhan2024pde}, we collect 50k pairs of boundary input $U(t)$ and output $Y(t)$ trajectory with safety labels based on safety constraints. The resolution of collected trajectories is consistent with the control frequency of each environment. 


\paragraph{Model training and evaluation metrics.} With the collected dataset from vanilla RL models,
we adopt the Fourier neural operator (FNO) \citep{li2020fourier} as the default neural operator model and train it with Markov neural operator (MNO) \citep{li2022learning} using the default hyper-parameters. For the neural BCBF training, following \cite{zhang2023exact, hu2024verification}, we use a 4-layer feedforward neural network with ReLU activations to parameterize BCBFs and incorporate \Cref{eq:loss_S} and \Cref{eq:total_loss} with default $\alpha=10^{-5}$ into the regular model training pipeline \citep{zhao2020synthesizing,dawson2022safe} to train time-dependent BCBF $\phi(t,Y)$ as default.  With the well-trained neural operator and neural BCBF, we solve the  QP of \Cref{eq:safety_filter} though CPLEX \citep{cplex} 
For the evaluation of safety filtering for RL controllers, we keep the original RL rewards from \cite{bhan2024pde} as a metric to show if the performance is compromised by the enhancement of safety constraints. Besides, we introduce two new metrics regarding boundary feasibility, \textit{Feasible Rate} and \textit{Average Feasible Steps}. \textit{Feasible Rate} is the ratio of trajectories that boundary feasibility in Definition \ref{def:pf} is achieved, i.e., the boundary output falls into the safe set and will not go out of it by the end of a single trajectory with finite steps. \textit{Average Feasible Steps} is the mean steps among boundary feasible trajectories in which the boundary output is consistently kept in the safe set until the end of the trajectory, characterizing how long the boundary feasibility is achieved and maintained.

\subsection{Results Comparison}
\label{sec:result_compare}
\paragraph{Comparison of vanilla models with safety filtering.}
From all three PDE environments in \Cref{tab:vanilla},  vanilla PPO and SAC with safety filtering outperform vanilla PPO and SAC in feasible rate and average feasible steps, demonstrating the effectiveness of safety filtering for boundary constraint satisfiability. Besides, the rewards in parabolic and hyperbolic equations can also be improved through filtering due to the alignment of boundary constraints and the stabilization goal. The reward of the filtered SAC model in the hyperbolic equation is compromised because the constraint $Y<0$ conflicts with the stabilization task of $Y\rightarrow 0$.
 In the 2D Navier-Stokes PDE, due to the inconsistency between the specific high-speed point boundary for constraint and the full 2D plane for reward, boundary feasibility is enhanced by safety filtering while rewards are compromised. 

\paragraph{Safety filtering performance based on constrained RL models.} To further show the plug-and-play efficacy of our safety filtering method, we present the filtering performance over the constrained RL models in \Cref{tab:safeRL} using the pre-trained BCBF, which is trained over data collected from vanilla RL models. We can see that compared to CPO \citep{achiam2017constrained}, the filtered controller tends to improve the boundary feasibility, especially for the stronger constraint $Y<0$. Safety filtering over SAC-Lag \citep{ha2020learning} will give higher feasibility rates over the boundary, while the average feasible steps slightly decrease because feasible steps along trajectories become more concentrated and less divergent after filtering. Besides, despite the potential conflict between boundary constraint and stabilization, the reward will not be hurt significantly via safety filtering.


\begin{table}[t]
\caption{Comparison of time-independent and time-dependent safety filtering in hyperbolic equations.}
\vspace{-7mm}
\begin{center}
\label{tab:ablation_different_BCBF}
\resizebox{0.85\textwidth}{!}{
    \begin{tabular}{cccc} \toprule
Different safety filtering              & \begin{tabular}[c]{@{}c@{}}Reward (mean$\pm$std)\\ (starting at $\sim$-300)\end{tabular}  & \begin{tabular}[c]{@{}c@{}}Feasible Rate\\ (100 episodes)\end{tabular}  &  \begin{tabular}[c]{@{}c@{}}Average Feasible Steps\\ ( 50 control steps)\end{tabular}  \\ \midrule
 PPO with filtering  of $\phi(Y)$& 162.3$\pm$44.5  & 0.63                        & 8.3\\
 PPO with filtering  of $\phi(t,Y)$ & {165.0}$\pm$43.7  & \textbf{0.71 }                       & \textbf{9.8}                                                          \\\midrule
 SAC with filtering of $\phi(Y)$  & 103.3$\pm$98.4 & 0.57                        & \textbf{15.7}       \\
 SAC with filtering of $\phi(t,Y)$& {103.4}$\pm$96.4 & \textbf{0.85}                        & 13.9       \\
\bottomrule
\end{tabular}
}
\end{center}
\vspace{-7mm}
\end{table}

\begin{table}[]
\caption{Filtering with  BCBF $\phi(t,Y)$ under different neural operators for 1D hyperbolic equation. }
\vspace{-7mm}
\begin{center}
\label{tab:FNO_MNO}
\resizebox{0.85\textwidth}{!}{
    \begin{tabular}{cccc} \toprule
Different neural operators                    & \begin{tabular}[c]{@{}c@{}}Reward (mean$\pm$std)\\ (starting at $\sim$-300)\end{tabular}  & \begin{tabular}[c]{@{}c@{}}Feasible Rate\\ (100 episodes)\end{tabular}  &  \begin{tabular}[c]{@{}c@{}}Average Feasible Steps\\ ( 50 control steps)\end{tabular}  \\ \midrule
PPO w. MNO \citep{li2022learning}     & 163.8$\pm$47.2  & \textbf{0.78}                        & 9.0                                                          \\
PPO w. FNO \citep{li2020fourier}   & {165.0}$\pm$43.7  & 0.71                        & \textbf{9.8}                                                         \\\midrule
SAC w. MNO \citep{li2022learning}     & 103.3$\pm$96.4  & 0.84                        & \textbf{14.7}                                                        \\
SAC w. FNO \citep{li2020fourier} & {103.4}$\pm$96.4 & \textbf{0.85}                        & 13.9       \\
\bottomrule
\end{tabular}
}
\end{center}
\vspace{-11mm}
\end{table}
\begin{figure}
\vspace{-2mm}
    \centering
    \includegraphics[width=0.75\linewidth]{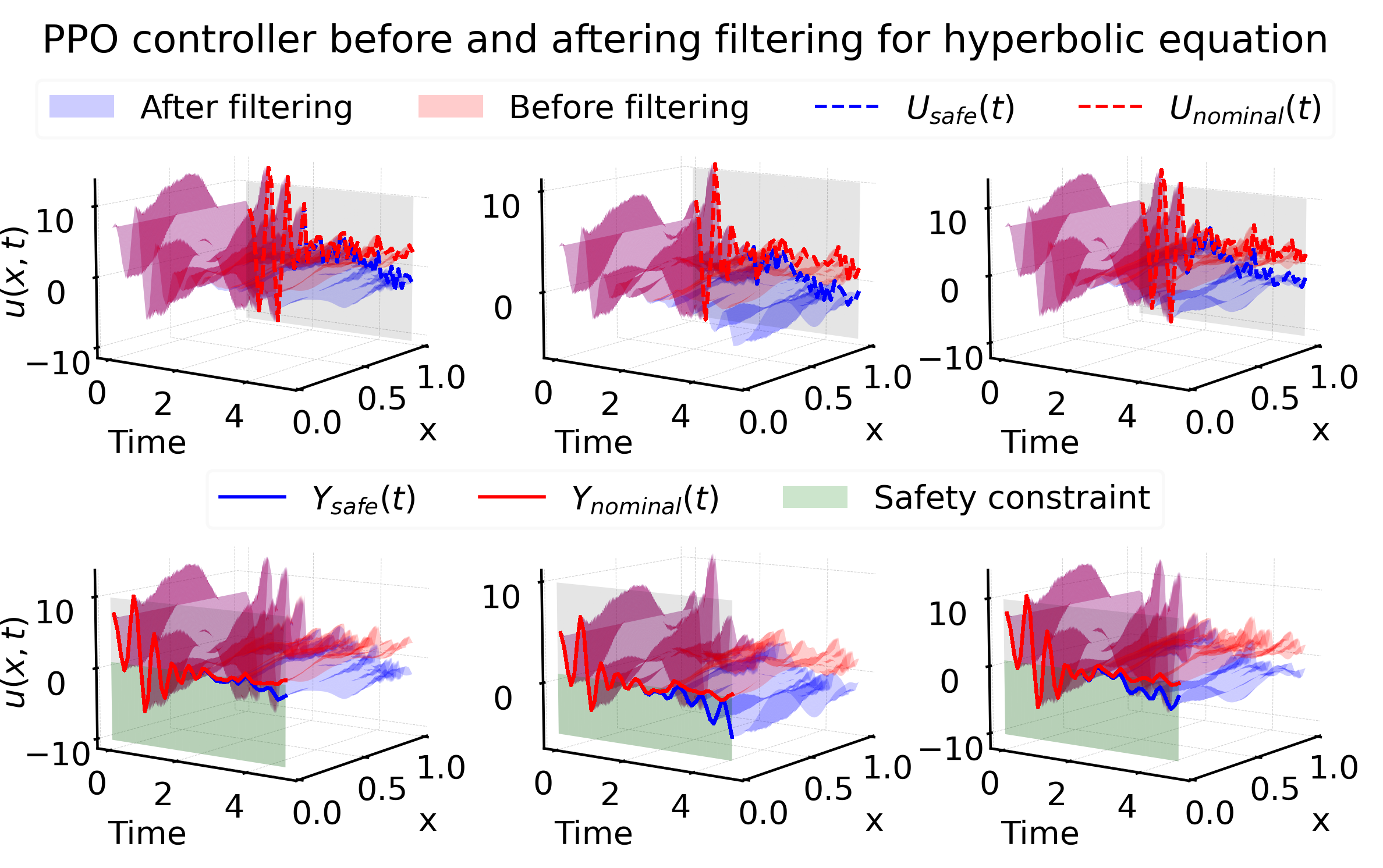}
    \vspace{-4mm}
    \caption{Visualization of three state trajectories $u(x,t)$ (left, mid, right) for hyperbolic equation under PPO controller \textcolor{blue}{with} and \textcolor{red}{without} safety  filtering. Boundary control inputs $U(t)$ are in dashed lines, and boundary outputs $Y(t)$ are in solid lines. The boundary  constraint $Y(t) < 1$ is in green.}
    \vspace{-6mm}
    \label{fig:visualize_ppo}
\end{figure}

\subsection{Ablation Study}
\label{sec:ablation}

\paragraph{Comparison of safety filtering using $\phi(Y)$ vs. $\phi(t,Y)$.}  
With different boundary control barrier functions in \Cref{tab:ablation_different_BCBF}, with the PPO model, safety filtering with $\phi(t,Y)$ outperforms filtering with $\phi(Y)$ in reward and boundary feasibility metrics, showing that time-dependent BCBF can distinguish the feasibility of the PDE boundary state more effectively by explicitly taking time as an input compared to the time-independent one. Based on the vanilla SAC model, 
reward and feasible rate with $\phi(t,Y)$ filtering is higher but the average feasible step is lower than $\phi(Y)$ filtering, because time-independent BCBF $\phi(Y)$ tends to have divergent performance with more non-feasible trajectories and more feasible steps for feasible trajectories.

\paragraph{Boundary mapping with different neural operators.} Here we compare two neural operators, FNO \citep{li2020fourier} and MNO \citep{li2022learning}, for learning the boundary mapping from control input $U(t)$ to output $Y(t)$ for 1D hyperbolic equation in \Cref{tab:FNO_MNO}. With the same time-dependent BCBF $\phi(t,Y)$, the safety filtering with FNO presents higher rewards under both PPO and SAC base models, showing that FNO is more suitable for learning low-resolution trajectories with 50 sampled points. Besides, MNO shows a better feasible rate and average feasible steps performance, especially with SAC as the base model, since the MNO model has a larger model complexity.

\paragraph{Qualitative visualization.} In this section, we visualize and compare multiple trajectories under the 1D hyperbolic equation using the PPO controller without and with safety filtering of $\phi(t,Y)$, as shown in \Cref{fig:visualize_ppo}. We can see that for each trajectory, the state value $u(x,t)$ after filtering is lower than that before filtering. More specifically, as time goes by, the filtered control input $U(t)_{\text{safe}}$ in blue dashed lines deviates more away from nominal control input $U(t)_{\text{nominal}}$ in red dashed lines, causing the filtered boundary output $Y(t)_{\text{safe}}$ in blue solid lines to satisfy the constraint $Y(t)<1$ compared to the nominal boundary output $Y(t)_{\text{nominal}}$ in red solid lines.

%% file: conclusion.tex
\section{Conclusion}
In this work, we introduce a novel safe PDE  boundary control framework using safety filtering with neural certification. A neural operator and a boundary control barrier function are learned from collected PDE boundary input and output trajectories within a given safe set.
We show that the change in the BCBF depends linearly on the change in the input boundary. Hence, safety filtering can be done by solving a quadratic programming problem to ensure the boundary feasibility.
Experiments on three challenging PDE control environments validate the effectiveness of the proposed method in terms of both general performance and constraint satisfaction.
